\documentclass[preprintnumbers,showpacs,floats,twocolumn,prd,aps]{revtex4}
\usepackage{amssymb,amsmath}
\usepackage{epsfig,hyperref}
\usepackage[svgnames]{xcolor}
\usepackage{pgf,tikz}
\usepackage{epstopdf}

\def\be{\begin{equation}}
\def\ee{\end{equation}}
\def\beq{\begin{eqnarray}}
\def\eeq{\end{eqnarray}}

\makeatletter
\renewcommand{\vec}[1]{\mbox{\boldmath$#1$}}
\newcommand{\arXiv}[2][]{\href{http://arxiv.org/abs/#2}{\texttt{arXiv:#2\@ifempty{#1}{}{ [#1]}}}}
\makeatother

\begin{document}

\title{Solar system tests of $\boldsymbol{f(R)}$ gravity}

\author{Jun-Qi Guo}%
\email{jga35@sfu.ca}
\affiliation{%
 Department of Physics, Simon Fraser University\\
8888 University Drive, Burnaby, BC Canada V5A 1S6
}%

\date{\today}

\begin{abstract}
In this paper, we revisit the solar system tests of $f(R)$ gravity. When the Sun sits in a vacuum, the field $f'$ is light, which leads to a metric different from the observations. We reobtain this result in a simpler way by directly focusing on the equations of motion for $f(R)$ gravity in the Jordan frame.
The discrepancy between the metric in the $f(R)$ gravity and the observations can be alleviated by the chameleon mechanism. The implications from the chameleon mechanism on the functional form $f(R)$ are discussed.
Considering the analogy of the solar system tests to the false vacuum decay problem, the effective potentials in different cases are also explored. The combination of analytic and numerical approaches enables us to ascertain whether an $f(R)$ model can pass the solar system tests or not.
\end{abstract}

\pacs{04.25.Nx, 04.50.Kd, 11.10.Lm, 95.36.+x, 98.80.Es, 98.80.Jk}
\maketitle

\section{Introduction \label{sec:introduction}}

The causes of the current cosmic acceleration have not yet been determined \cite{Riess:1998cb, Perlmutter:1998np, Riess:2004nr, Komatsu:2010fb, Planck:2013}, although various efforts have been made to identify it. In addition to considering the cosmological-constant and scalar-field approaches, the idea that an extended theory of gravity might account for the speed-up also seems to provide a straightforward explanation. In the Jordan frame (JF), one may replace the Ricci curvature scalar in the Einstein-Hilbert action with a function of the scalar,
\be S_{\text{JF}}=\frac{1}{16\pi G}\int d^{4}x \sqrt{-g}f(R)+S_M\left[g_{\mu\nu},\psi_{m}\right], \label{f_R_action} \ee
where $G$ is the Newtonian gravitational constant, and $\psi_{m}$ is the matter field [see Refs.~\cite{Sotiriou,Tsujikawa1} for reviews of $f(R)$ theory].

A viable $f(R)$ model should generate cosmic dynamics compatible with the cosmological observations~\cite{Amendola,Aviles,Guo}, and also pass the solar system tests, which place strong constraints on $f(R)$ gravity. The metric of the spherical spacetime for the Sun predicted by general relativity matches well with the observations. General relativity predicts that the parameterized post-Newtonian(PPN) parameter $\gamma$ is equal to 1, and the observational results show that $\gamma=1+O(10^{-4})$~\cite{Shapiro,Bertotti}. Therefore, general relativity should be recovered from $f(R)$ gravity in the solar system. However, if the Sun sits in a vacuum, the scalar field $f'$ will be very light, which will generate a metric considerably different from the observations~\cite{Sotiriou,Tsujikawa1,Chiba,Faulkner,Kobayashi,Kainulainen,Wayne_Hu,Pengjie_Zhang}. In this paper, we re-derive this result in a simpler way
by directly considering the equations of motion for $f(R)$ gravity in the JF. In addition, in PPN formalism, based on the empirical definition of the mass of the Sun, the PPN parameter $\alpha$ must be equal to 1. With this value, one obtains the formula for the deflection angle of a light pulse travelling around the Sun, and gets the PPN parameter $\gamma$ by comparing this formula with the observations. However, in $f(R)$ gravity, the PPN parameter $\alpha$ is generally not equal to 1. Therefore, a $\gamma$ can not be obtained by directly taking the ratio between the two modification terms in the $(t,t)$ and $(r,r)$ coefficients of the metric line element.

General relativity could be recovered through the chameleon mechanism~\cite{Sotiriou,Tsujikawa1,Faulkner,Justin1,Justin2,Navarro,Gu,Tsujikawa2,Tsujikawa3}. In this mechanism, the scalar field $f'$ is coupled to the matter density of the environment. $f'$ acquires a mass from the coupling, and then is suppressed by the mass, such that the $f(R)$ gravity can pass the solar system tests. The chameleon mechanism is usually implemented in the Einstein frame (EF), in which the condition on the existence of a thin shell is obtained in Refs.~\cite{Justin1, Justin2}. However, the matter density and the transformed scalar field are coupled in a complex way in the EF. Note that the $f(R)$ gravity is defined in the JF, and the picture is more intuitive in the JF, in which the coupling between the matter density and the scalar field $f'$ is much simpler than the form in the EF. In this paper, we discussed the chameleon mechanism in the JF, and also explored the implications from this mechanism on the functional form of $f(R)$.

In addition to the analytical method, numerical approach also provides an efficient way to study how the scalar field $f'$ behaves in the
effective potential. Taking the $R\ln R$ model and Hu-Sawicki model as examples, we run the numerical experiments solving the equation of motion
for $f'$ in different configurations. The results verify the arguments for the thin-shell condition obtained in the JF and the thin-wall approximation condition in the false-vacuum decay scenario, and further clarify whether an $f(R)$ model can avoid the solar system tests or not.

This paper is organized as follows.
In Sec.~\ref{sec:f_R gravity}, we introduce the framework of the solar system tests of $f(R)$ gravity.
In Sec.~\ref{sec:gamma_1_over_2}, a situation of the Sun sitting in a vacuum background is discussed.
In Sec.~\ref{sec:gamma_1}, working in the JF, we explore the chameleon mechanism, and its implications on the function form of $f(R)$.
In Sec.~\ref{sec:analogy}, the false-vacuum decay scenario is discussed.
In Sec.~\ref{sec:numerical}, numerical computations are performed to verify the thin-shell condition. In Sec.~\ref{sec:conclusions},
the results are summarized.

\section{Framework \label{sec:f_R gravity}}
\subsection[f(R) gravity]{$f(R)$ gravity}
A variation on the action for $f(R)$ gravity with respect to the metric yields gravitational equations of motion,
\be
f'R_{\mu\nu}-\frac{1}{2}f g_{\mu\nu} -\left(\nabla_{\mu}\nabla_{\nu}-g_{\mu\nu} \Box\right)f'= 8\pi GT_{\mu\nu},
\label{gravi_eq_fR} \ee
where $f'$ denotes the derivative of the function $f$ with respect to its argument $R$, and $\Box$ is the usual
notation for the covariant D'Alembert operator $\Box\equiv\nabla_{\alpha}\nabla^{\alpha}$. The dynamics of the degree of freedom $f'$ are determined
by the trace of Eq.~(\ref{gravi_eq_fR})
\be \Box f'=\frac{8\pi G}{3}T+\frac{1}{3}(2f-f'R),\label{trace_eq1}\ee
where $T$ is the trace of the stress-energy tensor $T_{\mu\nu}$. Identifying $f'$ by
\be \phi\equiv\frac{df}{dR}, \label{f_prime}\ee
and defining a potential $V(\phi)$ by
\be V'(\phi)\equiv\frac{dV}{d\phi}=\frac{1}{3}(2f-\phi R), \label{v_prime} \ee
one can rewrite Eq.~(\ref{trace_eq1}) as
\be \Box \phi=V'(\phi)+\frac{8\pi G}{3}T.\label{trace_eq2}\ee
The spherically symmetric coordinate system inside and outside of the Sun is set up as
\begin{eqnarray}
\frac{\phi}{r^{2}}(-1+B+rB') & = & -8 \pi G \rho-\frac{1}{2}(\phi R-f)\nonumber\\
 && -B\left[\phi''+\left(\frac{2}{r}+\frac{B'}{2B}\right)\phi'\right],
\label{tt_component}
\end{eqnarray}
\begin{eqnarray}
\frac{\phi}{r^{2}}\left(-1+B+rB\frac{N'}{N}\right) & = & 8\pi Gp-\frac{1}{2}(\phi R-f) \nonumber\\
  && -B\left(\frac{2}{r}+\frac{N'}{2N}\right)\phi ',
\label{rr_component}
\end{eqnarray}
where the prime $(')$ denotes the derivative with respect to $r$, $\rho$ is matter density, and $p$ is pressure.
The trace equation (\ref{trace_eq1}) in the coordinate system described by Eq.~(\ref{de_Sitter_metric}) reads
\be
B\left[\phi''+\left(\frac{2}{r}+\frac{B'}{2B}+\frac{N'}{2N}\right)\phi'\right]=\frac{8\pi
G}{3}(-\rho+3p)+V'(\phi).
\label{complete_trace_eq}\ee
When the gravitational field is weak and $\rho \gg p$, Eq.~(\ref{complete_trace_eq}) can be approximated as
\be \phi''+\frac{2}{r}\phi'=-\frac{8\pi G}{3}\rho+V'(\phi). \label{approx_trace_eq}\ee
The boundary conditions are
\be \lim_{r\to\infty}\phi(r)=\phi_{\infty}=\text{Const}, \mbox{ }\left. \frac{d\phi}{dr}\right|_{r=0}=0. \label{boundary_f_R}\ee
We will use the above formulas to explore the behaviors of $f(R)$ gravity in the solar system, and compare the theoretical results with the observations in PPN formalism.
\subsection{PPN formalism\label{sec:PPN_formalism}}
Generally, an isotropic coordinate is used in the PPN formalism~\cite{Chiba,Will,Berry}, while it is convenient to express the metric outside of the Sun in the spherical coordinate~\cite{Weinberg}
\begin{eqnarray}
ds^{2} & = & -\left[1-2\alpha\frac{GM}{r}+2(\beta-\alpha\gamma)\left(\frac{GM}{r}\right)^{2}\right]dt^2\nonumber\\
 && +\left[1+2\gamma\frac{GM}{r}\right]dr^{2}+r^{2}d\Omega^2,
\label{metric_spherical}
\end{eqnarray}
where $\alpha$, $\beta$, and $\gamma$ are PPN parameters, and $M$ is the solar mass. In general relativity, $\alpha=\beta=\gamma=1$, and the prediction of $\alpha=1$ follows from the empirical definition of the mass $M$. A centripetal acceleration for a slowly moving particle far from the center of the Sun is
\be -g=-\Gamma^{r}_{tt}=\frac{1}{2}\frac{\partial g_{tt}}{\partial r}=-\frac{\alpha GM}{r^{2}},\ee
and the solar mass is measured by setting $g$ to be equal to $MG/r^{2}$. Therefore, we must set $\alpha$ to be equal to 1~\cite{Weinberg}. Then we can rewrite Eq.~(\ref{metric_spherical}), to first order of $MG/r$, as
\be ds^{2} = -\left(1-\frac{2GM}{r}\right)dt^2 +\left(1+\gamma\frac{2GM}{r}\right)dr^{2}+r^{2}d\Omega^2.
\label{metric_spherical_2}
\ee
A light pulse will be deflected by the Sun when it travels around the Sun. Denote $d$ as the closest distance between the light pulse and the center of the Sun. Then, to first order of $MG/d$, the deflection angle for the metric Eq.~(\ref{metric_spherical_2}) can be expressed as~\cite{Weinberg}
\be \delta \theta = \left( \frac{1+\gamma}{2} \right)\frac{4M}{d}, \label{deflection_PPN}\ee
which is the same as the one derived in isotropic coordinate~\cite{Will}. In general relativity, $\gamma$ is equal to 1 and then the deflection angle $\delta \theta_{\text{GR}}$ is equal to $4M/d$. The results from VLBI (very-long-baseline radio interferometry) observations show that $\gamma-1=(-1.7\pm 4.5)\times 10^{-4}$ \cite{Shapiro,Will_2}.
\section{Assume the Sun sits in a vacuum \label{sec:gamma_1_over_2}}
\subsection{The metric}

In $f(R)$ gravity, if we assume that the Sun sits in a vacuum, the metric outside of the Sun, described by Eq.~(\ref{metric_f_R}), is dramatically different from the observations. In Refs.~\cite{Chiba,Sotiriou,Faulkner,Kobayashi,Kainulainen,Wayne_Hu,Pengjie_Zhang}
the metric Eq.~(\ref{metric_f_R}) is obtained by a perturbation method or by transforming $f(R)$ theory from the JF into the EF. In this paper, we will re-compute the metric in a simpler way by directly focusing on the equations of motion in the JF.

We assume that the Sun is sitting in a vacuum, and that in Eq.~(\ref{approx_trace_eq}) $V'(\phi)$ is negligible in comparison to
$\phi''$ and $2\phi'/r$. Note that at infinity, $\phi$ is expected to be close to the de Sitter value $\phi_0$, for which $V'(\phi)|_{\phi=\phi_0}=0$. Thus,
the solution to Eq.~(\ref{approx_trace_eq}), inside and outside of the Sun, can be written as
\be \phi(r)|_{\text{interior}} \approx \phi_0 +\frac{\epsilon}{2r_{0}}\left[3-\left(\frac{r}{r_0}\right)^{2}\right],
\label{phi_interior}\ee
\be \phi(r)|_{\text{exterior}} \approx \phi_0 +\frac{\epsilon}{r},
\label{phi_exterior}\ee
where $\epsilon=2GM/3$, $M$ is the solar mass, and $r_0$ is the radius of the Sun. When both $(\phi R-f)/2$ and $p$ can be neglected,
the equations of motion (\ref{tt_component}) and (\ref{rr_component}) become
\be \frac{1}{r^{2}}(-1+B+rB') \approx -\frac{2}{3\phi_0} 8 \pi G \rho, \label{B_prime} \ee
\be \frac{N'}{N}\approx\frac{\frac{\phi}{r^{2}}(1-B)-\frac{2B}{r}\phi'}{B\left(\frac{\phi}{r}
+\frac{\phi'}{2}\right)}. \label{N_prime} \ee
Notice that Eq.~(\ref{B_prime}) differs from the corresponding equation in general relativity only by a factor of $2/(3\phi_0)$.
This means that the solutions of $B(r)$ inside and outside of the Sun can be written as
\be B|_{\text{interior}}\approx 1-\frac{2\epsilon_1}{r_0}\left(\frac{r}{r_0}\right)^{2},
\label{B_interior} \ee
\be B(r)|_{\text{exterior}}\approx 1-\frac{2\epsilon_1}{r}, \label{B_exterior} \ee
where $\epsilon_1=2GM/(3\phi_0)$. Substituting of Eqs.~(\ref{phi_interior})(\ref{B_interior}) and Eqs.~(\ref{phi_exterior})(\ref{B_exterior})
into Eq.~(\ref{N_prime}) yields the solutions for $N(r)$ inside and outside of the Sun, respectively,
\be N(r)|_{\text{interior}}\approx \exp\left[\frac{4\epsilon_1}{r_0}\left(\frac{r}{r_0}\right)^{2}+C_{1}\right], \label{N_interior0} \ee
\be N(r)|_{\text{exterior}}\approx C_2\left(1-\frac{4\epsilon_1}{r}\right). \label{N_exterior0} \ee
Letting $C_2$ equal to 1 and requiring $N(r)$ to be continuous at $r$ equal to $r_{0}$, we have
\be N(r)|_{\text{exterior}} \approx 1-\frac{4\epsilon_1}{r}, \label{N_exterior} \ee
\be N(r)|_{\text{interior}} \approx 1-\frac{8\epsilon_1}{r_0}+\frac{4\epsilon_1}{r_0}\left(\frac{r}{r_0}\right)^{2}. \label{N_interior} \ee
In summary,
\be
\phi(r)\approx \left\{
\begin{array}{l l}
  \phi_0\left\{1+\frac{\epsilon_1}{2r_{0}}\left[3-\left(\frac{r}{r_0}\right)^{2}\right]\right\} & \quad \mbox{for $r<r_{0}$}, \\
   \\
  \phi_0\left(1+\frac{\epsilon_1}{r}\right) & \quad \mbox{for $r>r_{0}$},\\
\end{array} \right.
\label{phi_analytic}
\ee
\be
B(r)\approx \left\{
\begin{array}{l l}
  1-\frac{2\epsilon_1}{r_0}\left(\frac{r}{r_0}\right)^{2} & \quad \mbox{for $r<r_{0}$}, \\
   \\
  1-\frac{2\epsilon_1}{r} & \quad \mbox{for $r>r_{0}$},\\
\end{array} \right.
\label{B_analytic}
\ee
\be
N(r)\approx \left\{
\begin{array}{l l}
  1-\frac{8\epsilon_1}{r_0}+\frac{4\epsilon_1}{r_0}\left(\frac{r}{r_0}\right)^{2} & \quad \mbox{for $r<r_{0}$}, \\
   \\
  1-\frac{4\epsilon_1}{r} & \quad \mbox{for $r>r_{0}$}.\\
\end{array} \right.
\label{N_analytic}
\ee
For comparison, we also list the corresponding quantities in general relativity: $\phi(r)$ is always equal to $1$ and
\be
B_{\text{GR}}(r)\approx \left\{
\begin{array}{l l}
  1-\frac{3\epsilon}{r_0}\left(\frac{r}{r_0}\right)^{2} & \quad \mbox{for $r<r_{0}$}, \\
   \\
  1-\frac{3\epsilon}{r} & \quad \mbox{for $r>r_{0}$},\\
\end{array} \right.
\ee
\be
N_{\text{GR}}(r)\approx \left\{
\begin{array}{l l}
  1-\frac{6\epsilon}{r_0}+\frac{3\epsilon}{r_0}\left(\frac{r}{r_0}\right)^{2} & \quad \mbox{for $r<r_{0}$}, \\
   \\
  1-\frac{3\epsilon}{r} & \quad \mbox{for $r>r_{0}$}.\\
\end{array} \right.
\ee

More generally, when the matter density and the radius of the Sun are large enough for some $f(R)$ models, $\phi(r)$ could be almost constant
at $0<r<r_1<r_0$, therefore $V'(\phi)-8\pi G\rho/3 \approx 0$. Note that in this case the chameleon mechanism is functioning.
At $r_1<r<r_0$, the field $\phi(r)$ varies with $r$, $\phi''+2\phi'/r=-8\pi G\rho/3+V'(\phi)$.
In this case, the solutions of $B(r)$ and $N(r)$ are a bit more complicated. However, the forms of $\phi(r)$, $B(r)$, and $N(r)$ will not change outside of the Sun, except that in the definition of $\epsilon_1$, $M$ is replaced by
\be M_{\text{eff}}\approx4\pi \int_{r_{1}}^{r_0} \left[\rho - \frac{3V'(\phi)}{8\pi G}\right]r^{2} dr.\ee
\subsection{Compare the theoretical results with the observations}
Substitution of Eqs.~(\ref{B_exterior}) and (\ref{N_exterior}) into Eq.~(\ref{de_Sitter_metric}) yields
\begin{subequations}\label{metric_f_R}
\begin{align}
ds^{2}&=-\left(1-\frac{8}{3\phi_0}\frac{GM}{r}\right)dt^2 +\left(1+\frac{4}{3\phi_0}\frac{GM}{r}\right)dr^{2}+r^{2}d\Omega^2, \label{metric_f_R_1}\\
&=-\left(1-\frac{2GM_0}{r}\right)dt^2+ \left(1+\frac{GM_0}{r}\right)dr^{2}+r^{2}d\Omega^{2}, \label{metric_f_R_2}
\end{align}
\end{subequations}
where $M_0=4M/(3\phi_0)$. Comparison of Eqs.~(\ref{metric_spherical_2}) and (\ref{metric_f_R_1}) shows that, in $f(R)$ gravity, $\alpha=4/(3\phi_0)$ and $\gamma=2/(3\phi_0)$. The de Sitter value $\phi_0$ may not be equal to 3/4. Consequently, $M_0$ may not be equal to $M$ as in general relativity, and $\alpha(=4/(3\phi_0))$ may not be equal to 1, which violates the requirement that $\alpha$ must be equal to 1 as argued in Sec.~\ref{sec:PPN_formalism}. Therefore, it may not be appropriate to directly take the ratio between the two terms of $4GM/(3\phi_0)$ and $8GM/(3\phi_0)$ in Eq.~(\ref{metric_f_R_1}) to be the value of the PPN parameter $\gamma$, as done in general relativity case, and claim that $\gamma$ is equal to $1/2$. Another instructive approach to confront $f(R)$ gravity with the observations is to compute the deflection angle of a light pulse travelling around the Sun. Compare Eq.~(\ref{metric_spherical_2}) with Eq.~(\ref{metric_f_R_2}) and
use Eq.~(\ref{deflection_PPN}), one obtains the deflection angle for a light pulse in $f(R)$ gravity
\be
\delta \theta_{f(R)} = \left( \frac{1+\gamma}{2} \right)\frac{4M_0}{d} = \frac{1}{\phi_0}\frac{4M}{d}
=\frac{\delta \theta _{\text{GR}}}{\phi_0}.
\label{deflection_PPN_f_R}
\ee
Equation (\ref{deflection_PPN_f_R}) shows that the deflection angle in $f(R)$ gravity happens to differ from the one in general relativity by a factor of $\phi_0$, which generally is not equal to $1$ in $f(R)$ gravity. The differences of the PPN parameter $\alpha's$ and the deflection angels between general relativity and $f(R)$ gravity can be explained as follows. Compared to general relativity, in $f(R)$ gravity, there is one more degree of freedom $f'$. Equivalently, an extra force is operating. This force affects the metric and then the trajectory of the light pulse. In the next section, we will discuss the chameleon mechanism, in which the degree of freedom $f'$ is suppressed and therefore the $f(R)$ gravity may avoid the solar system tests.

In Refs.~\cite{Berry,Clifton}, a PPN parameter $\gamma$ equal to 1 is obtained via a linear perturbation of the metric for $f(R)$ gravity in flat Minkowski space. In some $f(R)$ models, e.g. $f(R)=R+\alpha R^{2}$ where $\alpha$ is a parameter, the de Sitter curvature (obtained from $V'(\phi)=0$) is zero and hence the background space can be Minkowski space. However, in the dark-energy-oriented $f(R)$ gravity, the curvature of the background is not equal to zero, but equal to the cosmological constant. A linearized analysis in the de Sitter space will lead to the metric described by Eq.~(\ref{metric_f_R})~\cite{Sotiriou,Chiba}.

\section{The chameleon mechanism \label{sec:gamma_1}}
%
\subsection{The chameleon mechanism in the EF}
The discrepancy between the theoretical results and the observations in solar system tests of $f(R)$ gravity could be avoided through the chameleon mechanism~\cite{Sotiriou,Tsujikawa1,Faulkner,Justin1,Justin2,Navarro,Gu,Tsujikawa2,Tsujikawa3}. The $f(R)$ gravity
can be transformed into the EF. The field $\phi$ is re-scaled to $\tilde{\phi}=\sqrt{3/2}m_{\text{pl}}\ln \phi$, where $m_{\text{pl}}$ is the Planck mass
and $8\pi G=m_{\text{pl}}^{-2}$. Consider that the Sun sits in the solar system background, which has a non-zero matter density. The new field $\tilde{\phi}$ can acquire a mass from its coupling to the matter density of the environment both inside and outside of the Sun. The field $\tilde{\phi}$ is suppressed by this mass. In order to avoid the solar system tests, the field $\tilde{\phi}$ should be suppressed to satisfy~\cite{Faulkner}
\be \frac{|\tilde{\phi}^{\infty}_{\text{min}}-\tilde{\phi}^{c}_{\text{min}}|}{\Phi_c}\frac{\sqrt{3/2}}{m_{\text{pl}}}
\leq 3.5\times10^{-5},
\label{thin_shell}\ee
where $\Phi_c\approx 10^{-6}$ is the Newtonian potential at the solar surface.

\subsection{The chameleon mechanism in the JF\label{sec:JF_gamma_1}}
\begin{figure}[!htbp]
\epsfig{file=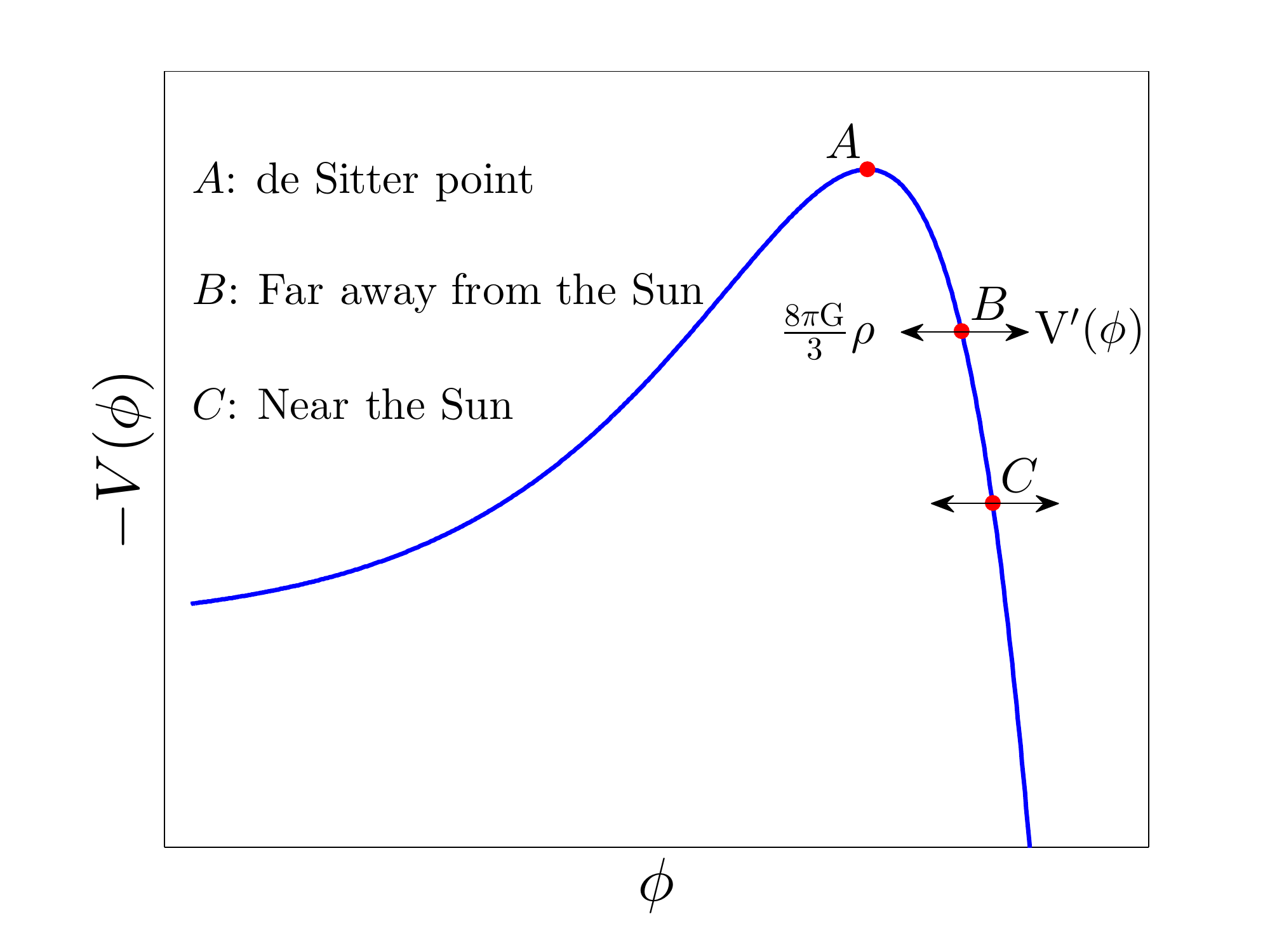, width=6.6cm}\\
\setlength{\abovecaptionskip}{-0.1cm}
\caption{Description of the chameleon mechanism in the Jordan frame.}
\label{fig:negative_potential}
\end{figure}

We present the chameleon mechanism in the JF in this subsection, which has a simpler format than the one in the EF. General relativity predicts that $\gamma$ is equal to 1, which matches the observations very well. By comparing the equations of motion in $f(R)$ gravity, (\ref{tt_component}) and (\ref{rr_component}), with those in general relativity,
\be \frac{1}{r^{2}}(-1+B+rB')= -8 \pi G \rho, \label{GR_tt_component}\ee
\be \frac{1}{r^{2}}\left(-1+B+rB\frac{N'}{N}\right)=8\pi Gp,\label{GR_rr_component}\ee
one can see that in order to obtain a $\gamma$ equal to 1 in $f(R)$ gravity, $f(R)$ gravity should be reduced to general relativity in the solar system. The corresponding matter density ranges from $\rho \approx \rho_{\mbox{\tiny{Sun}}}\sim \mbox{g}/\mbox{cm}^3$ to
$\rho \approx \rho_{\mbox{\tiny{solar-system}}}\sim 10^{-8}\mbox{g}/\mbox{cm}^3$,
\be f(R) \approx  R, \mbox{ and } f' \approx 1.\label{GR_recovery}\ee
Therefore, with Eq.~(\ref{approx_trace_eq}), which is the equation of motion for $\phi$, we obtain that from inside the Sun to the places far away from the Sun the following equation should be satisfied
\be V'(\phi)-\frac{8\pi G}{3}\rho\approx 0. \label{v_rho_1} \ee
Taking the spatial coordinate $r$ in Eq.~(\ref{approx_trace_eq}) as the \lq\lq time\rq\rq~coordinate,
\be \ddot{\phi}+\frac{2}{r}\dot{\phi}=-\left[-V'(\phi)+\frac{8\pi G}{3}\rho\right].\ee
Then, as shown in Fig.~\ref{fig:negative_potential}, Points $B$ and $C$ correspond to the respective quasi-stationary states for the field $\phi$ outside and inside of the Sun, and $|\phi_B-\phi_C|\ll 1$. However, if the matter density of the environment is zero, the field $\phi$ will move to Point $A$,
and a metric Eq.~(\ref{metric_f_R}) different from the observations will be obtained.

When $\phi'\rightarrow0$ and $\phi''\rightarrow0$, Eqs.~(\ref{tt_component}) and (\ref{rr_component}) become
\be \frac{\phi}{r^{2}}(-1+B+rB')= -8\pi G\rho-\frac{1}{2}(\phi R-f),
\label{tt_component3}\ee
\be \frac{\phi}{r^{2}}\left(-1+B+rB\frac{N'}{N}\right)=8\pi G p-\frac{1}{2}(\phi R-f). \label{rr_component3}\ee
Letting $\eta_1=8\pi G\rho + (\phi R-f)/2$ and $\eta_2=-8\pi G p+(\phi R-f)/2$, we obtain
\be B(r) \approx 1+\frac{C_1}{r}-\frac{\eta_1}{3}r^{2}, \ee
\be N(r) \approx C_2 \exp\left[-\int \frac{C_1/r^2+(\eta_2-\eta_{1}/3)r}{1+C_1/r-\eta_1 r^{2}/3}dr\right]. \ee
When $\eta_{1}r\ll C_1/r^2$ and $\eta_{2}r\ll C_1/r^2$, one obtains
\be B(r)\approx N(r)\approx1+\frac{C_1}{r}, \ee
and $\gamma \approx 1$.
\subsection{Requirements on the $f(R)$ format \label{sec:requirements_on_f_R}}
The requirements on the fields $\tilde{\phi}$ and $\phi$ to pass the solar system tests have been discussed in the above two subsections. In this subsection, we study the implications from these requriements on the form of the function $f(R)$.

Consider a small variation of Eq.~(\ref{v_rho_1})
\be V''(\phi)\cdot \delta\phi \approx \frac{8\pi G}{3}\delta\rho. \label{variation_v_rho}\ee
As discussed in the above subsection, when general relativity is restored, we have
$R\approx 8\pi G\rho_m$ and $|\delta\phi|\ll 1$. These together with Eq.~(\ref{variation_v_rho}) imply that
\be |\delta\phi|\approx\frac{8\pi G}{3}\frac{|\delta\rho|}{|V''(\phi)|}\sim \frac{R}{|V''(\phi)|} \ll 1
\label{v_dub_prime_matter_domi}.\ee
With Eq.~(\ref{v_prime}) defining $V'(\phi)$, we have
\be V''(\phi)=\frac{f'-f''R}{3f''}. \label{v_dubprime} \ee
Substitution Eq.~(\ref{v_dubprime}) into Eq.~(\ref{v_dub_prime_matter_domi}) yields that
\be f'\gg f''R. \label{condition_on_f_R} \ee
Equation~(\ref{condition_on_f_R}) can be interpreted as follows. Write the function $f(R)$ as
\be f(R)=R+A(R),\ee
where $R$ is the main term, and $A(R)$ is the modification term. Then Eqs.~(\ref{GR_recovery}) and (\ref{condition_on_f_R}) imply that the corrections should be smaller than the main terms at three orders of derivative:
\be |A(R)|\ll R, \mbox{ } |A'(R)|\ll 1, \mbox{ and } A''(R)\ll \frac{1}{R}. \label{condition_on_f_R_2} \ee
If general relativity is restored inside the Sun, which means that $|A(R)|\ll R$ and $|A'(R)|\ll 1$. To guarantee that general relativity recovery is valid from inside the Sun to places far away from the Sun, $A'(R)$ should also change considerably slowly with respect to $R$, and Eq.~(\ref{condition_on_f_R}) shows that the change should be slower than $1/R$.

In addition to the solar system tests, the $f(R)$ gravity should also be reduced to general relativity in the early universe so that the large-scale structure can be formed. Therefore, the recovery requirements on the $f(R)$ format in the two cases are essentially same. The derivation of Eq.~(\ref{condition_on_f_R_2}) is very similar to the corresponding one in the cosmological evolution aspect, see Ref.~\cite{Guo}. On the other side, the $f(R)$ gravity should deviate from general relativity at low curvature scale, so that a cosmic acceleration can be generate in the late universe. Therefore, the parameters in $f(R)$ models should take appropriate values to balance the requirements from both high and low curvature scales.
%
\section{False vacuum decay and solar system tests of $f(R)$ gravity\label{sec:analogy}}
From the mathematical point of view, the solar system tests of $f(R)$ gravity are very similar to the false vacuum decay discussed in Ref.~\cite{Coleman}.
The conclusions in Ref.~\cite{Coleman} provide a pictorial description to the thin shell problem in $f(R)$ gravity. In this section, we first briefly introduce the scenario of false vacuum decay, then discuss solar system tests of $f(R)$ gravity using the arguments of this scenario.

\subsection{False vacuum decay}

\begin{figure*}
  \begin{tabular}{cc}
  \hspace{0pt}\epsfig{file=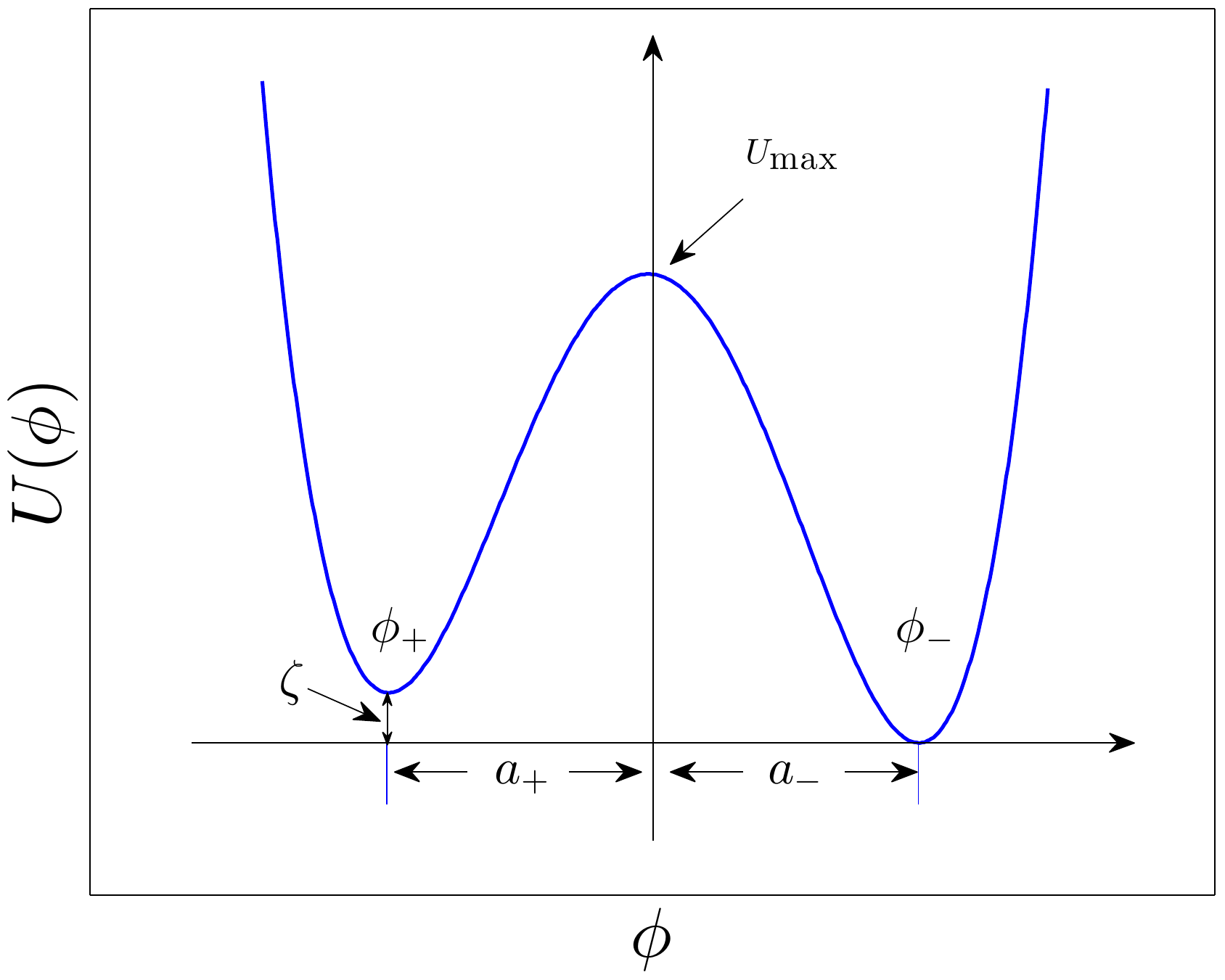, width=5.4cm} &
  \hspace{30pt}\epsfig{file=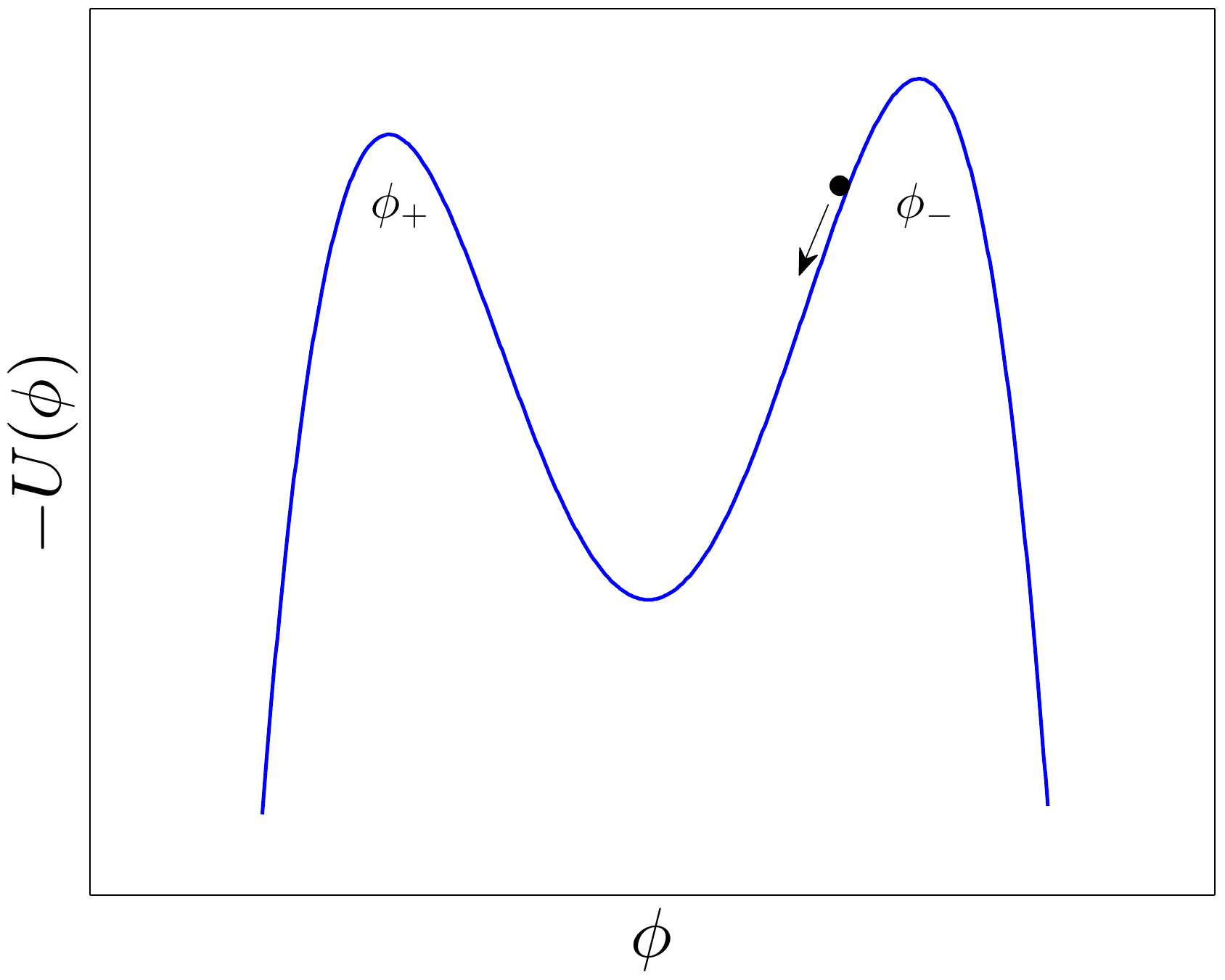, width=5.4cm} \\
  \hspace{7pt}  (a) &
  \hspace{39pt} (b) \\
  \end{tabular}
\caption{Instanton in the false vacuum decay. (a) Pictorial description of the thin-shell condition expressed by Eq.~(\ref{thin_shell_condition_new_3}).
  (b) The inverted potential $-U(\phi)$. In the beginning, the field $\phi$ quasi-statically stays at the right maximum of the inverted effective potential $-U(\phi)$ for a long \lq\lq time\rq\rq~due to the large friction force $2\dot{\phi}/r$. Then, after a long \lq\lq time\rq\rq~(if $r$ is large enough), the friction force becomes negligible, the field $\phi$ will run through the valley instantly and then slowly approach the left maximum of $-U(\phi)$.}
  \label{fig:U_decay}
\end{figure*}

Consider a single scalar field in four-dimensional spacetime with the nonderivative interactions
\be \mathcal{L}=\frac{1}{2}\partial_{\mu}\phi\partial^{\mu}\phi-U(\phi). \ee
Let $U$ possess two relative minima, $\phi_{\pm}$, only $\phi_{-}$ is an absolute minimum as shown in Fig.~\ref{fig:U_decay}. Assume the energy difference between the two minima, $\zeta$, is tiny, and denote $U_{\mbox{max}}$ as the local maximum of $U$ in the first order of $\zeta$. If $\phi$ stays at the minimum $\phi_{+}$ initially, quantum effects can make $\phi$ penetrate the barrier and approach to $\phi_{-}$. Thus, $\phi_{+}$ is called a false vacuum. The Euclidian (imaginary-time) equation of motion for $\phi$ is
\be \left(\frac{\partial^{2}}{\partial \tau ^{2}}+\nabla^{2}\right)\phi=U'(\phi),\label{equation_decay_1}\ee
where $\tau=it$, and the prime denotes differentiation with respect to $\phi$.
Define $\rho=(\tau^2+|\vec{x}|^2)^{1/2}$. Then, in the three-dimensional spherical symmetry, Eq.~(\ref{equation_decay_1}) becomes
\be \frac{d^{2}\phi}{d\rho ^{2}}+\frac{3}{\rho}\frac{d\phi}{d\rho }=U'(\phi).\label{equation_decay_2}\ee
The boundary conditions are set as
\be \lim_{\rho\to\infty}\phi(\rho)=\phi_{+}, \mbox{ } \left. \frac{d\phi}{d\rho}\right|_{\rho=0}=0. \label{boundary_decay_4}\ee
Equation~(\ref{equation_decay_2}) can be studied in the language of classical mechanics. Take $\rho$ as \lq\lq time\rq\rq~ variable, the field $\phi$ moves in $-U(\phi)$ as shown in Fig.~\ref{fig:U_decay}.b. Assume $\phi$ stays very close to $\phi_{-}$ initially. Due to friction, $\phi$ will remain close to $\phi_{-}$ until some very long time after. The field will run through the valley quickly, then approach $\phi_{+}$ very slowly. The place where $\phi$ moves fast is called thin-wall. The condition for the validity of the thin-wall approximation can be equivalently expressed in the following three formats \cite{Coleman}
\be \frac{\zeta}{8U_{\text{max}}} \ll 1  \Longleftrightarrow \frac{\Delta a}{a} \ll 1 \Longleftrightarrow \frac{\Delta \left(\mu^{2}\right)}{\mu^{2}} \ll 1. \label{thin_shell_condition_new_3} \ee
With the illustration of Fig.~\ref{fig:U_decay},
$\Delta a = |a_{+} - a_{-}|$,
$a=(a_{+} + a_{-})/2$,
$\mu^{2}=\left[U''(\phi_{+})+U''(\phi_{-})\right]/2$,
and $\Delta \left(\mu^{2}\right)=|U''(\phi_{+})-U''(\phi_{-})|$. Next, we will use these results to examine solar system tests of $f(R)$ gravity.

\subsection{Solar system tests of $f(R)$ gravity in context of instantons}
It is possible to rewrite Eq.~(\ref{approx_trace_eq}) as
\be \phi''+\frac{2}{r}\phi'=V_{\text{eff}}'(\phi), \label{approx_trace_eq_2}\ee
where
\be V_{\text{eff}}'(\phi)= -\frac{8\pi G}{3}\rho+V'(\phi). \label{V_eff_f_R}\ee
The boundary conditions are described by Eq.~(\ref{boundary_f_R}). The dynamics of $\phi(r)$ defined by Eqs.~(\ref{approx_trace_eq})-(\ref{boundary_f_R}) are almost identical to the scenario of false vacuum decay, except that the spacetime has one dimension less in the former than in the latter.

For convenience, one may take the spatial coordinate as the \lq\lq time\rq\rq ~coordinate, and rewrite Eq.~(\ref{approx_trace_eq_2}) in a \lq\lq dynamical\rq\rq~format as is done in last subsection
\be \ddot{\phi}+\frac{2}{r}\dot{\phi}=-\left[-V_{\text{eff}}'(\phi)\right], \label{approx_trace_eq_3}\ee
where the overdot still denotes the derivative with respect to the spacial coordinate $r$. The field $\phi$ is required to satisfy the boundary conditions described by Eq.~(\ref{boundary_f_R}). For $f(R)$ models satisfying the conditions expressed by Eq.~(\ref{condition_on_f_R_2}), general relativity can be well restored when the matter density is much greater than the cosmological constant. In this case, the field $\phi(\equiv f')$ is always very close to $1$ from inside to places far away from the Sun, and has a fast transition between the values inside and outside the Sun. Consequently, the two maxima of $-V_{\text{eff}}$ for these models are almost at the same \lq\lq height\rq\rq. In the beginning, the field $\phi$ quasi-statically stays at the right maximum of the inverted potential $-V_{\text{eff}}$ for a long \lq\lq time\rq\rq~due to the large friction force $2\dot{\phi}/r$. Then, after a long \lq\lq time\rq\rq~(when $r$ is large enough), the friction force becomes negligible such that the field $\phi$ will run through the valley instantly and then slowly approach the left maximum of $-V_{\text{eff}}$. In this case, $\phi$ is an instanton, and has a thin shell near the Sun's surface in the solar system tests problem.
However, for $f(R)$ models not well satisfying the condition expressed by Eq.~(\ref{condition_on_f_R_2}), general relativity may be recovered only for a very short period of matter density, and the field $\phi$ can deviate significantly from $1$ from inside the Sun to places far away from the Sun. Then, a slow transition for the field $\phi$ between the values inside and outside of the Sun will take place. In this case, the two maxima of the effective potential can be quite different. Both the instanton and non-instanton cases will be numerically implemented in the rest of this paper.
\section{Numerical computations of $f(R)$ models \label{sec:numerical}}
In order to verify the analytic arguments in Sections~\ref{sec:gamma_1} to \ref{sec:analogy}, we numerically investigate the profile of $\phi(r)$ for a sphere (including the Sun) sitting in a background with non-zero matter density. When the function $f(R)$ can trace the Ricci scalar $R$ closely enough in the solar system, the field $\phi$ will be very close to $1$, and has a fast but tiny drop near the solar surface. The two maxima of the effective potential will be almost same high, and the field $\phi$ is an instanton across the valley of the effective potential. However, if the $f(R)$ function deviates a lot from the Ricci scalar $R$ in the solar system, the field $\phi$ will change significantly from $1$ and has a long-term transition between the values inside and outside of the Sun. Consequently, the heights of the two maxima of the effective potential can be very different, and the corresponding $f(R)$ model has difficulty to pass the solar system tests. To justify these results, in this section, we will numerically explore the behaviors of $\phi(r)$ in the solar system taking the $R\ln R$ model and the Hu-Sawicki model as examples.

\subsection{The $R\ln R$ model}
The $f(R)$ gravity can be generated from a renormalization group flow. When the running of the gravitational coupling
is defined by a beta-function similar to the (quantum) function in quantum chromodynamics, one is led to the $R\ln R$ model~\cite{Frolov},
\be f(R)=R\left(1+\alpha_0\ln\frac{R}{R_{0}}\right), \label{f_R_logR} \ee
where $\alpha_0$ and $R_{0}$ are positive constants. The de Sitter curvature
\be \Lambda\equiv R_{0}e^{-1/\alpha_{0}+1},\label{dS_curvature_RlnR}\ee
is exponentially suppressed in comparison to $R_0$. The cosmic dynamics of this model are explored in our previous work~\cite{Frolov,Guo}.

For this model, the function $f(R)$ can be rewritten as
\be f(R)=R(\phi-\alpha_0), \label{f_R_logR_2}\ee
with
\be \phi\equiv f'=1+\alpha_0+\alpha_0 \ln\frac{R}{R_0}.\label{f_prime_RlnR}\ee
When $\phi''$ and $2\phi'/r$ are negligible compared to $V'(\phi)$ and $8\pi G\rho/3$ in Eq.~(\ref{approx_trace_eq}), one obtains
\be \phi \approx 2\alpha_0 + \alpha_0 W(X), \label{phi_RlnR} \ee
where $X=8\pi G\rho/\Lambda$ and $W(X)$ is the Lambert $W$ function. Equations~(\ref{f_R_logR_2}) and (\ref{phi_RlnR}) show that, at general-relativistic limit, $\phi\approx 1$, $R\sim R_0$, and
\be \alpha_0 \approx \frac{1}{W(X)} \ll 1. \ee
$W(X)\approx \ln(X)$ when $X\gg 1$. Then, with Eq.~(\ref{phi_RlnR}), $\phi$ is logarithmically dependent on $X$ when $X \gg 1$.
This model is therefore reduced to general relativity only for a certain regime of curvature or matter density even at high curvature scale. This is quite different from some other models, such as the Hu-Sawicki model~\cite{Wayne_Hu}, the Starobinsky model~\cite{Starobinsky}, and the exponential model ~\cite{Linder,Bamba,Elizalde}, in which $f(R)$ goes to general relativity once the matter density $\rho$ is above a certain value.

Regarding the low curvature regime, in order for the $f(R)$ gravity to generate a cosmic acceleration in the late universe, the de Sitter curvature and hence $\alpha_0$ can not be too small, see Eq.~(\ref{dS_curvature_RlnR}). Consequently, an appropriate value for $\alpha_0$ needs to be chosen to reconcile the tension between the requirements at the high and low curvature scales. Under the reconciliation, the running of $\phi$ makes it hard for the $R\ln R$ model to pass the solar system tests, as will be discussed below.

Utilizing Newton's iteration method, we numerically solve Eq.~(\ref{approx_trace_eq}) to obtain $\phi(r)$ for a sphere (which can be the Sun) in a background with non-zero matter density in different configurations. As far as the units are concerned, in this paper, we set the parameters to be dimensionless. We let the radius of the Sun $r_0$ equal to 1, in which case the densities of the Sun, the solar system and the dark energy are $2.1\times 10^{-6}$, $6\times10^{-14}$ and $10^{-34}$, respectively. We describe the matter density profile around the sphere approximately as
\be \rho = \frac{\rho_{\text{\tiny{sphere}}}}{ 1+e^{10(r-r_0)} } + \rho_{\text{\tiny{background}}}. \ee
For simplicity, we set $R_0$ equal to 1. We consider the following three cases in sequence: i) a thin shell of $\phi(r)$ exists, ii) a thick shell of $\phi(r)$ exists, iii) solar case, where the field $\phi$ does not sits at one minimum of $V_{\text{eff}}(\phi)$ inside the Sun, and a shell does not exist.

\begin{figure*}[t!]
  \begin{tabular}{ccc}
    \hspace{-3pt}\epsfig{file=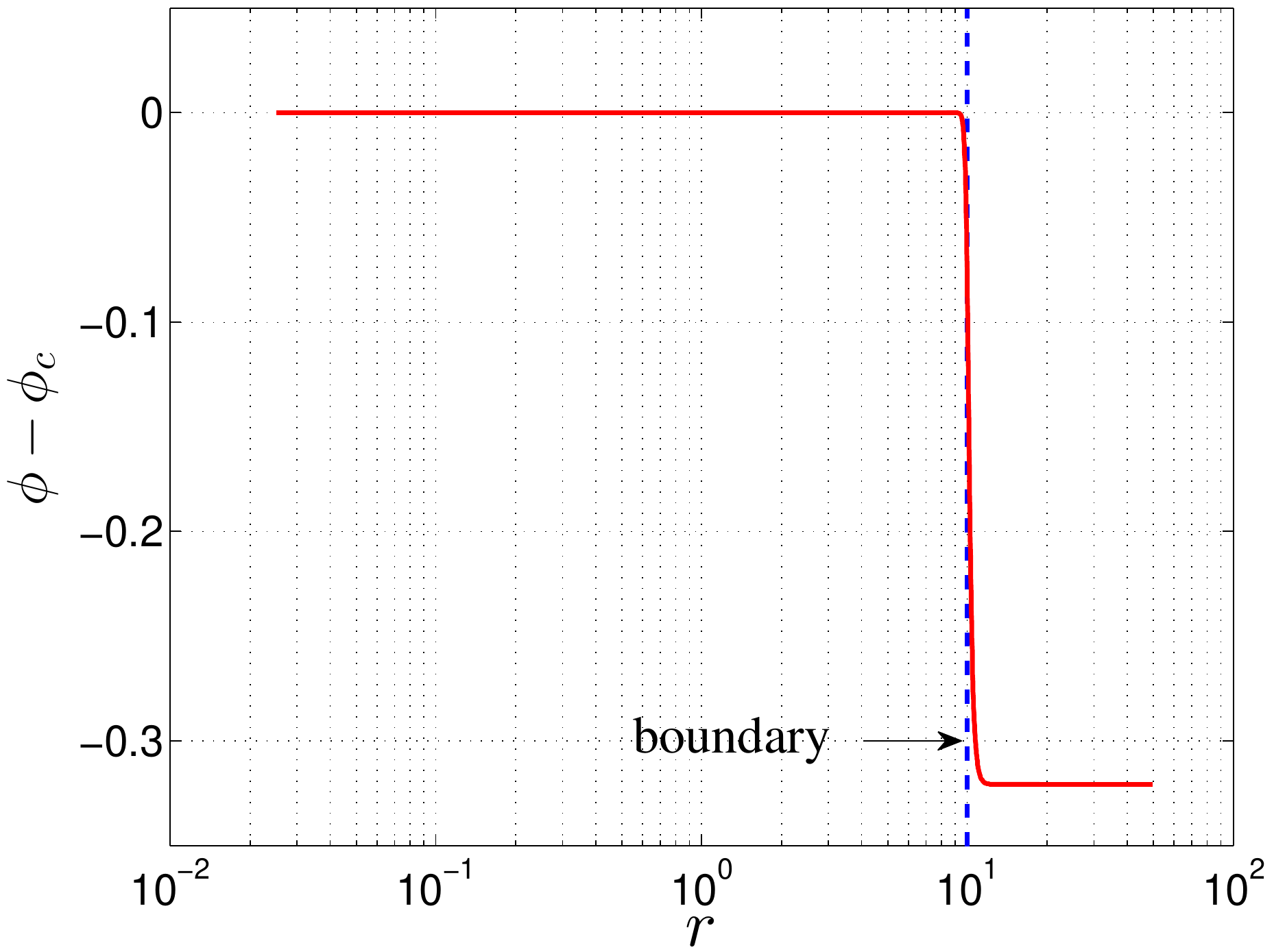, width=6cm} &
    \hspace{19pt}\epsfig{file=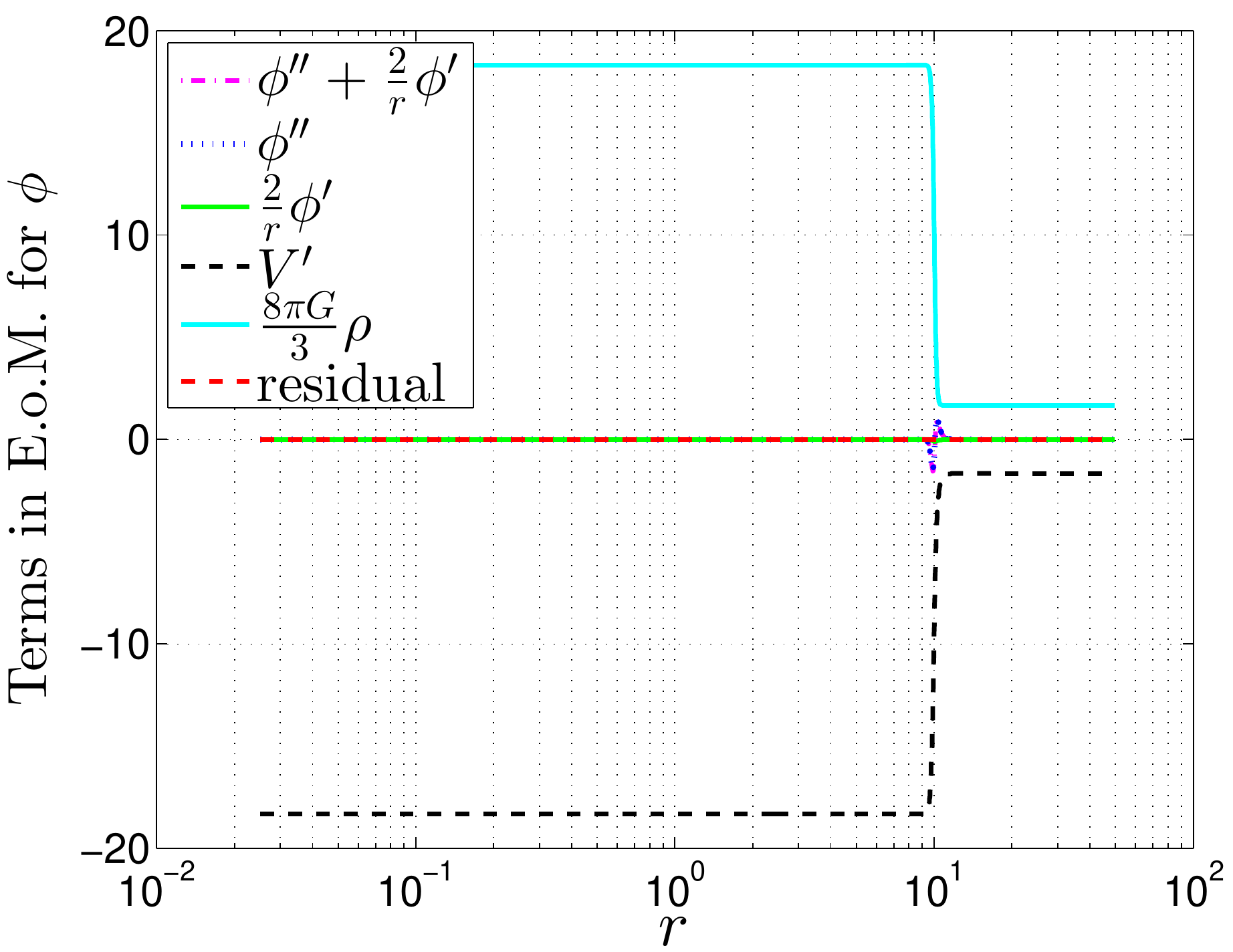, width=6cm} \\
    \hspace{14pt}(a) &
    \hspace{35pt}(b) \\
    \epsfig{file=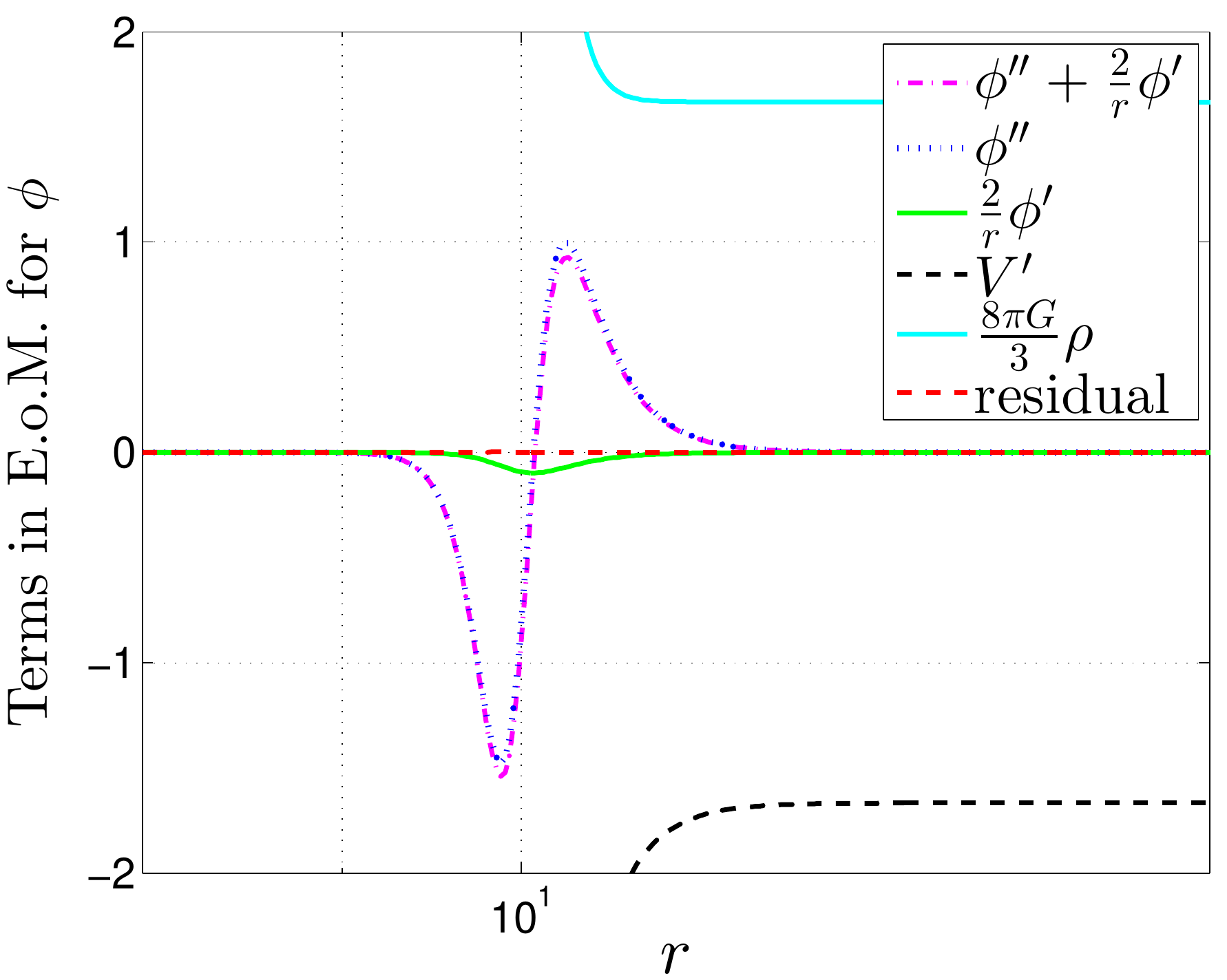, width=5.7cm} &
    \hspace{12pt}\epsfig{file=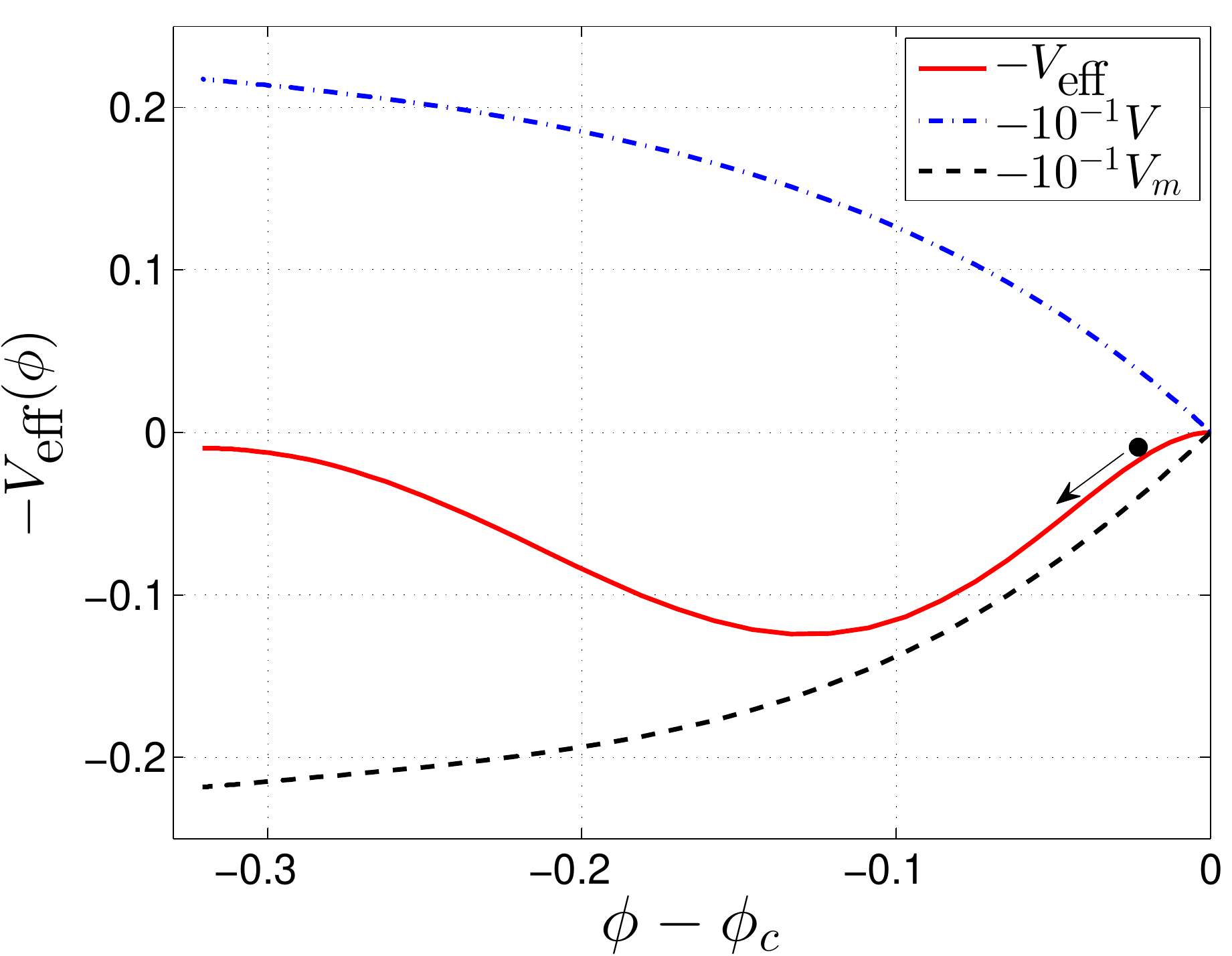, width=6cm,height=4.55cm} \\
    \hspace{8pt}(c) Zoom-in of (b)&
    \hspace{30pt}(d) \\
  \end{tabular}
\caption{(Color online) Numerical solution for the $R\ln R$ model when the parameters are set up to make the field $\phi$ heavy.
  (a) The field $\phi(r)$. $\phi(r)$ is coupled to the matter density from inside to outside of the sphere, and a thin shell exists at the surface of the sphere. $\phi_c(=1.70)$ is the value of $\phi$ at the center of the sphere.
  (b) The terms in the equation of motion (E.o.M.) for $\phi$ (\ref{approx_trace_eq}). From inside to outside of the sphere, Eq.~(\ref{approx_trace_eq}) is approximated as $V'(\phi)\approx 8\pi G\rho/3$. The absolute value of the friction force $|2\phi'/r|$ is much less than that of the net force $|\phi''|$, which allows the field $\phi$ to cross the valley of $-V_{\text{eff}}$ instantly.
  (c) Zoom-in of (b) near the surface of the sphere.
  (d) The inverted effective potential $-V_\text{eff}(\phi)$. The two maxima of $-V_\text{eff}(\phi)$ are almost at the same height, and $\phi(r)$ is an instanton in $-V_\text{eff}(\phi)$. $V_{\text{eff}}=V+V_m$.
  }
  \label{fig:solution_RlnR_set2}
\end{figure*}

Generally a thin-shell solution of $\phi(r)$ could exist, when i) the matter densities of the sphere and the background are high, ii) the gap between the two matter densities is not too large, iii) the sphere is large enough, iv) and the parameters take appropriate values so that the $f(R)$ model does not deviate much for general relativity at the curvature scale above the one of the background. Take $\alpha_0=0.015$ and $R_0=1$. Using the same units as the ones in the solar system case, we set the matter densities of the sphere and the background at $55$ and $5$, respectively. The radius of the sphere $r_0$ is $10$. The results for this set of parameters are shown in Fig.~\ref{fig:solution_RlnR_set2}. A thin-shell solution for $\phi$ exists at the surface of the sphere, see Fig.~\ref{fig:solution_RlnR_set2}.a. In this configuration, $\phi$ stays at the coupling state, for which $V'(\phi)\approx 8\pi G\rho/3$, from inside to outside of the sphere, and the absolute value of the friction force $|2\phi'/r|$ is much less than that of the net force $|\phi''|$, as shown in Fig.~\ref{fig:solution_RlnR_set2}.b and c. Consequently, $\phi$ can instantly cross the valley of $-V_{\text{eff}}(\phi)$ and, as plotted in Fig.~\ref{fig:solution_RlnR_set2}.c, the two maxima of the inverted effective potential $-V_{\text{eff}}$ are almost at the same height, and $\phi$ is an instanton in $-V_{\text{eff}}$. The potentials $V$, $V_{\text{eff}}$, and $V_m$ in Fig.~\ref{fig:solution_RlnR_set2}.c are defined by Eqs.~(\ref{v_prime}), (\ref{V_eff_f_R}), and $V'_m=-8\pi G\rho/3$, respectively. Equation~(\ref{V_eff_f_R}) implies that $V_{\text{eff}}=V+V_m$.

\begin{figure*}[t!]
  \begin{tabular}{ccc}
    \hspace{-14pt}\epsfig{file=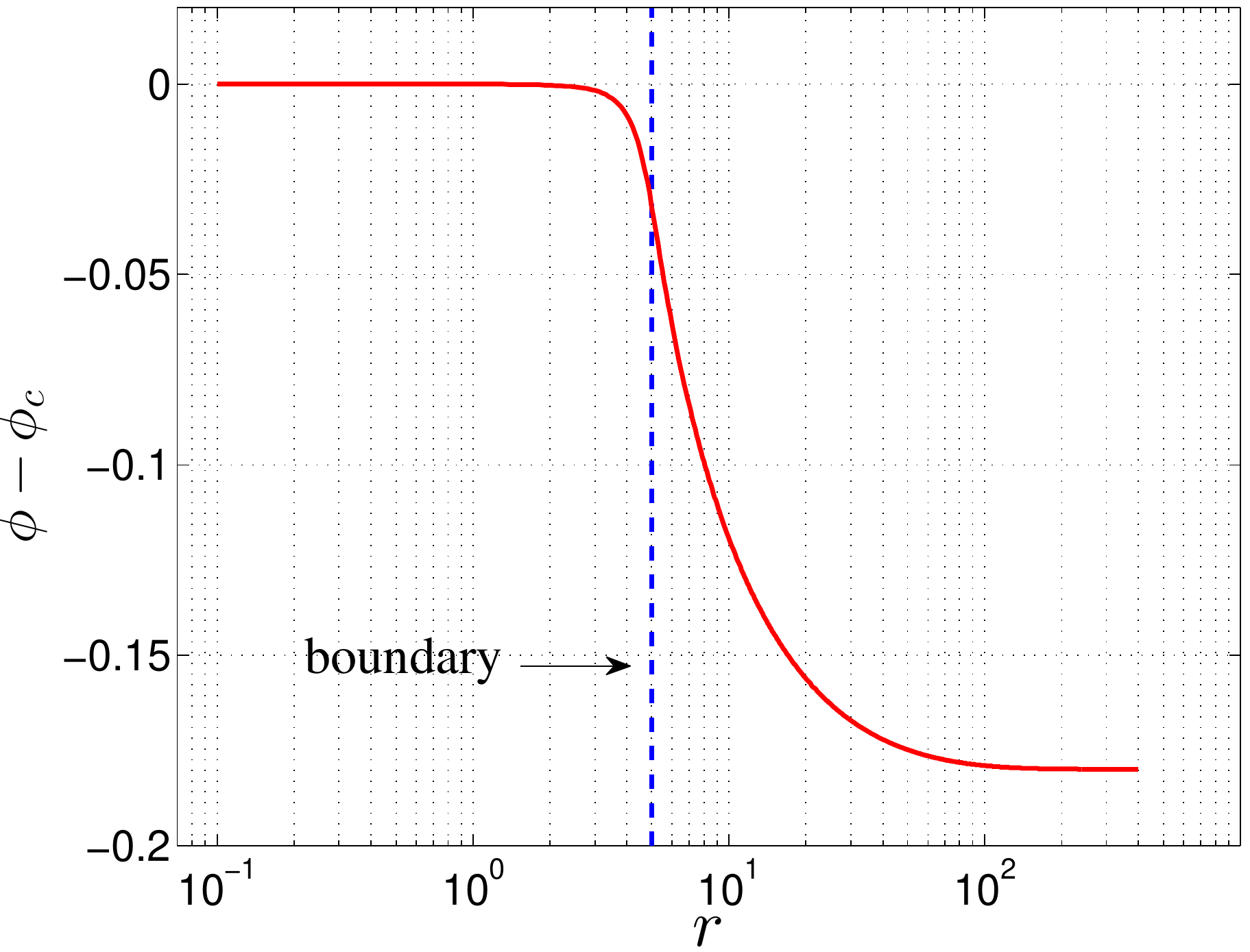, width=5.9cm} &
    \hspace{20pt}\epsfig{file=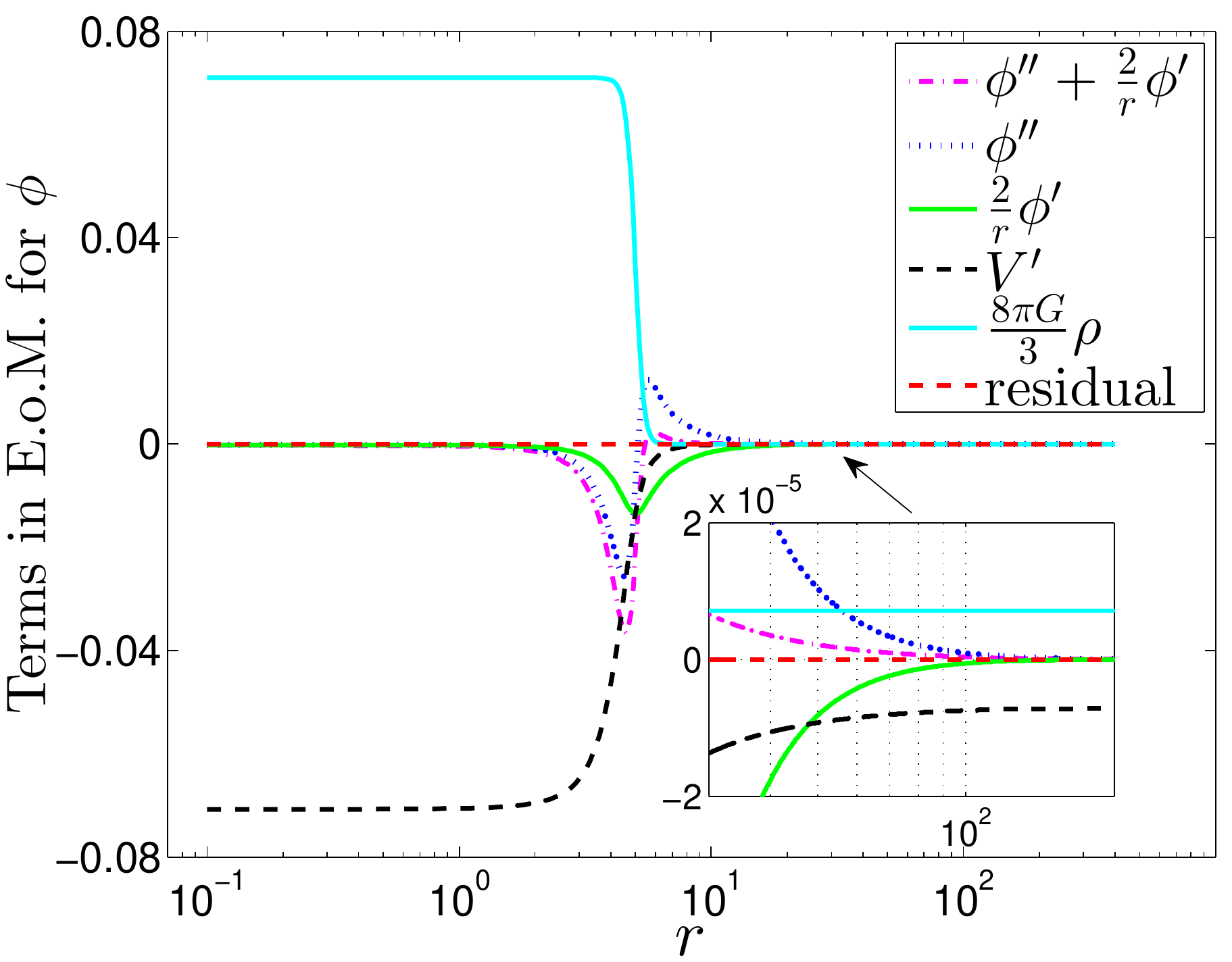, width=5.9cm} \\
    \hspace{8pt}(a) &
    \hspace{43pt}(b) \\
  \hspace{-7pt}\epsfig{file=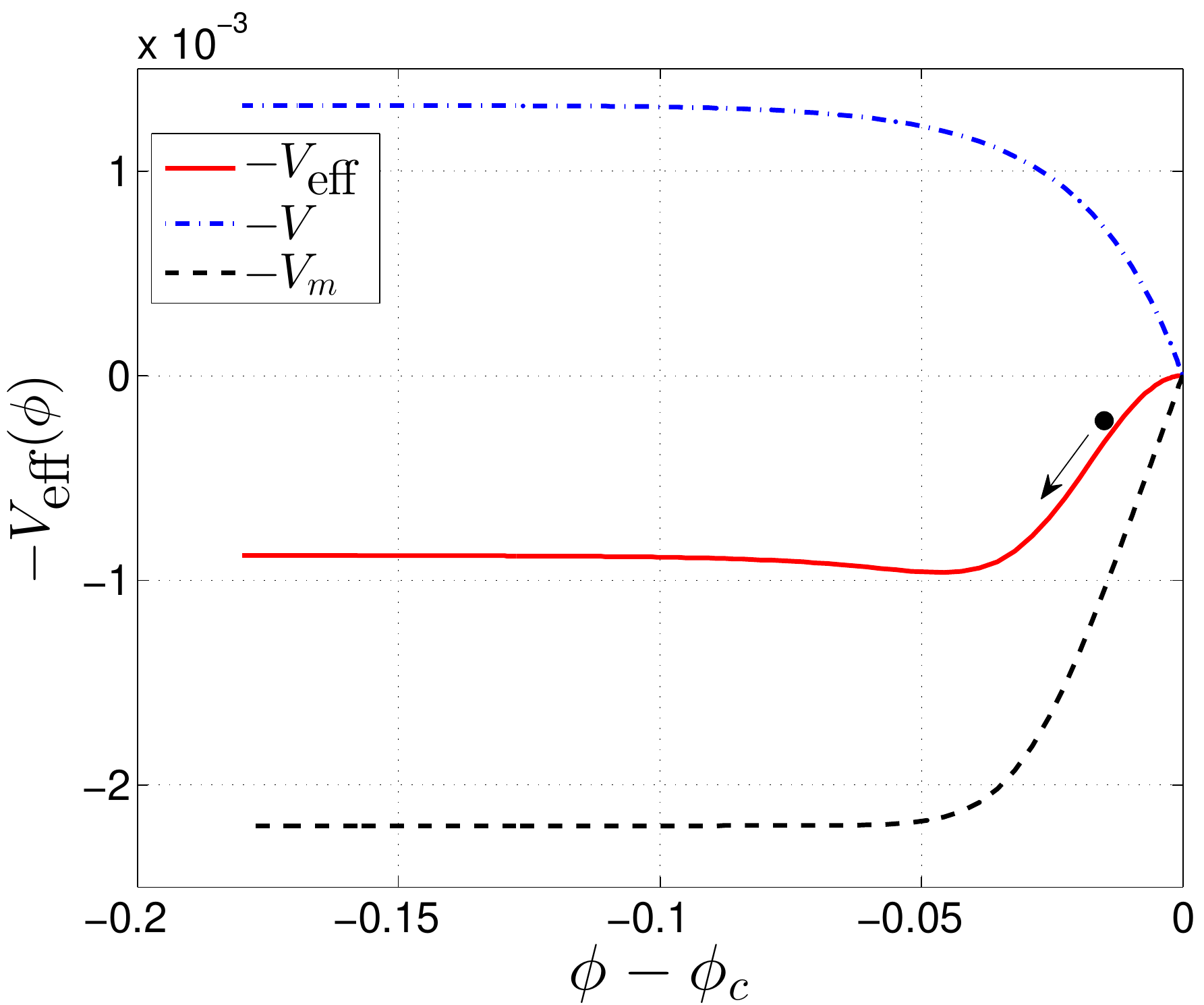, width=5.8cm,} \\
    \hspace{6pt}(c) \\
  \end{tabular}
  \caption{(Color online) Numerical solution for the $R\ln R$ model in an intermediate case.
  (a) The field $\phi(r)$. At places well inside and far away from the sphere, the field $\phi$ is coupled to the matter density of the environment. Near the surface of the sphere, the field $\phi$ is decoupled from the matter density and has a slow rather than instant roll (in other words, $\phi$ has a thick shell). $\phi_{c}(=0.99)$ is the value of $\phi$ at the center of the sphere.
  (b) The terms in the equation of motion for $\phi$ (\ref{approx_trace_eq}). At places well inside and far away from the sphere, Eq.~(\ref{approx_trace_eq}) becomes $V'(\phi)-8\pi G\rho/3\approx 0$. Just outside of the sphere, Eq.~(\ref{approx_trace_eq}) is approximated as $\phi''+2\phi'/r\approx 0$.
  (c) The inverted effective potential $-V_\text{eff}(\phi)$. The left maximum of $-V_\text{eff}(\phi)$ is lower than the right one. A large fraction of potential energy of the field $\phi$ is exhausted by the friction force in the rolling-down precess. $V_{\text{eff}}=V+V_m$.
  }
  \label{fig:solution_RlnR_set1}
\end{figure*}

Now we consider a more challenging configuration. Take $\alpha_0=0.02$ and $R_0=1$. The matter densities of the sphere and the background are $2.1\times10^{-1}$ and $2.1\times10^{-5}$, respectively. The radius of the sphere $r_0$ is 5. The results for this set of parameters are shown in Fig.~\ref{fig:solution_RlnR_set1}. As shown in Figs.~\ref{fig:solution_RlnR_set1}.a and b, in this situation, at places well inside and far away from the sphere, the field $\phi$ stays at the coupling state, and the equation of motion for $\phi$ (\ref{approx_trace_eq}) becomes $V'(\phi)-8\pi G\rho/3\approx 0$.  Near the surface the sphere, Eq.~(\ref{approx_trace_eq}) is approximated as $\phi''+2\phi'/r\approx 0$. In this case, $\phi$ has a thick shell. In the mean time, although $-V_{\text{eff}}(\phi)$ still has two maxima, they are not at the same height, as shown in Fig.~\ref{fig:solution_RlnR_set1}.c.

\begin{figure*}[t!]
  \begin{tabular}{ccc}
    \epsfig{file=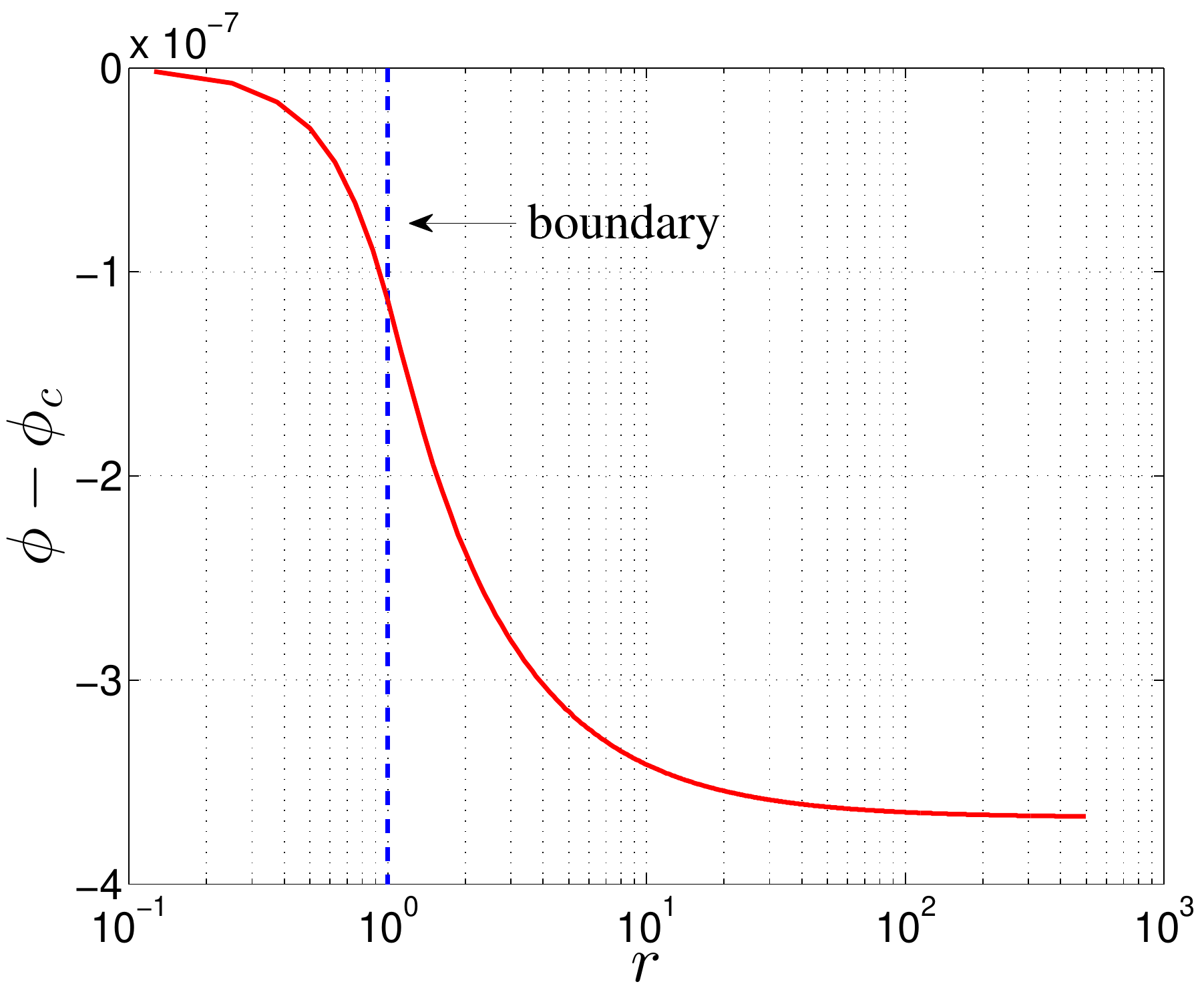, width=5.8cm} &
    \hspace{20pt}\epsfig{file=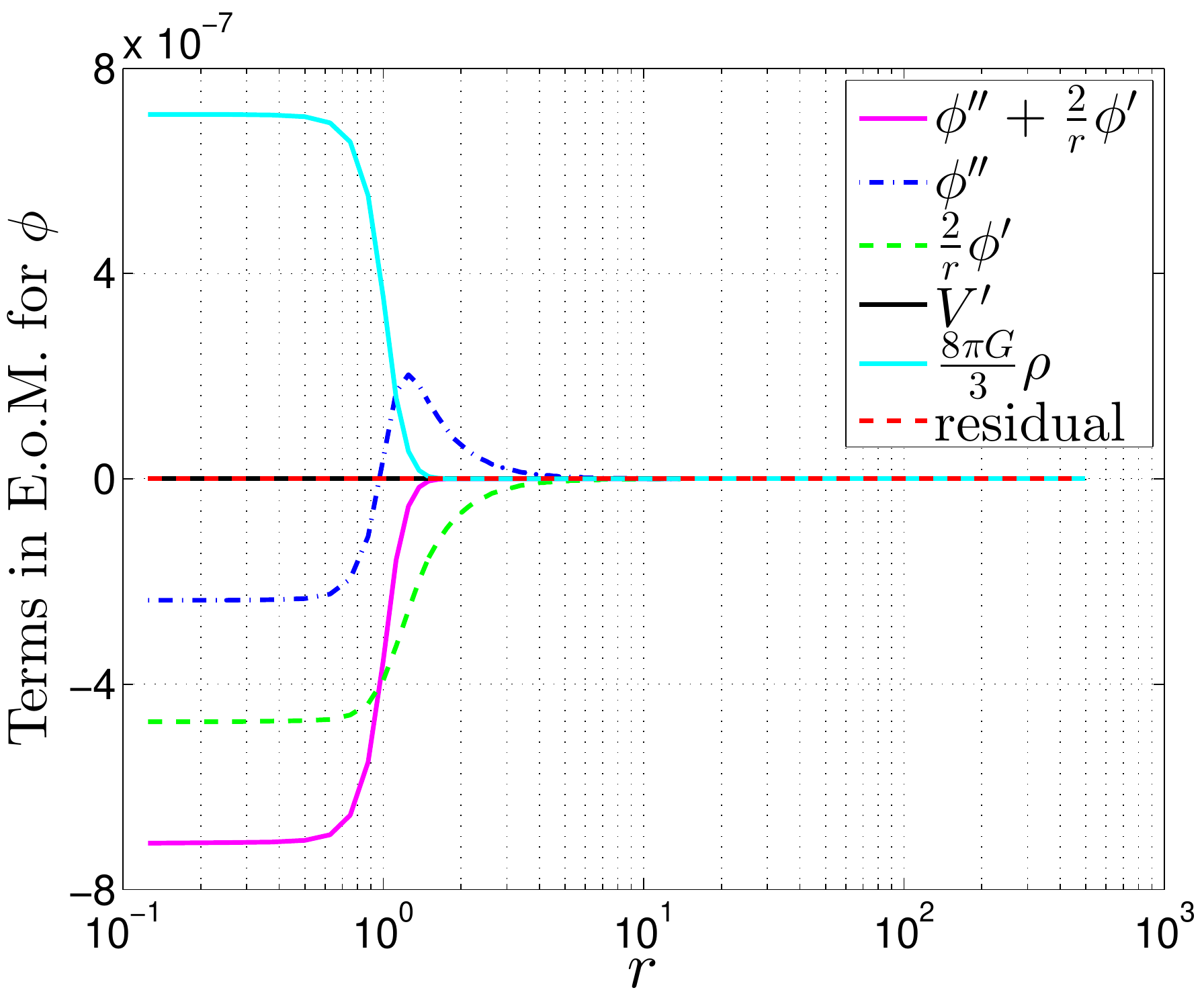, width=5.8cm} \\
    \hspace{13pt}(a) &
    \hspace{32pt}(b) \\
    \hspace{-8pt}\epsfig{file=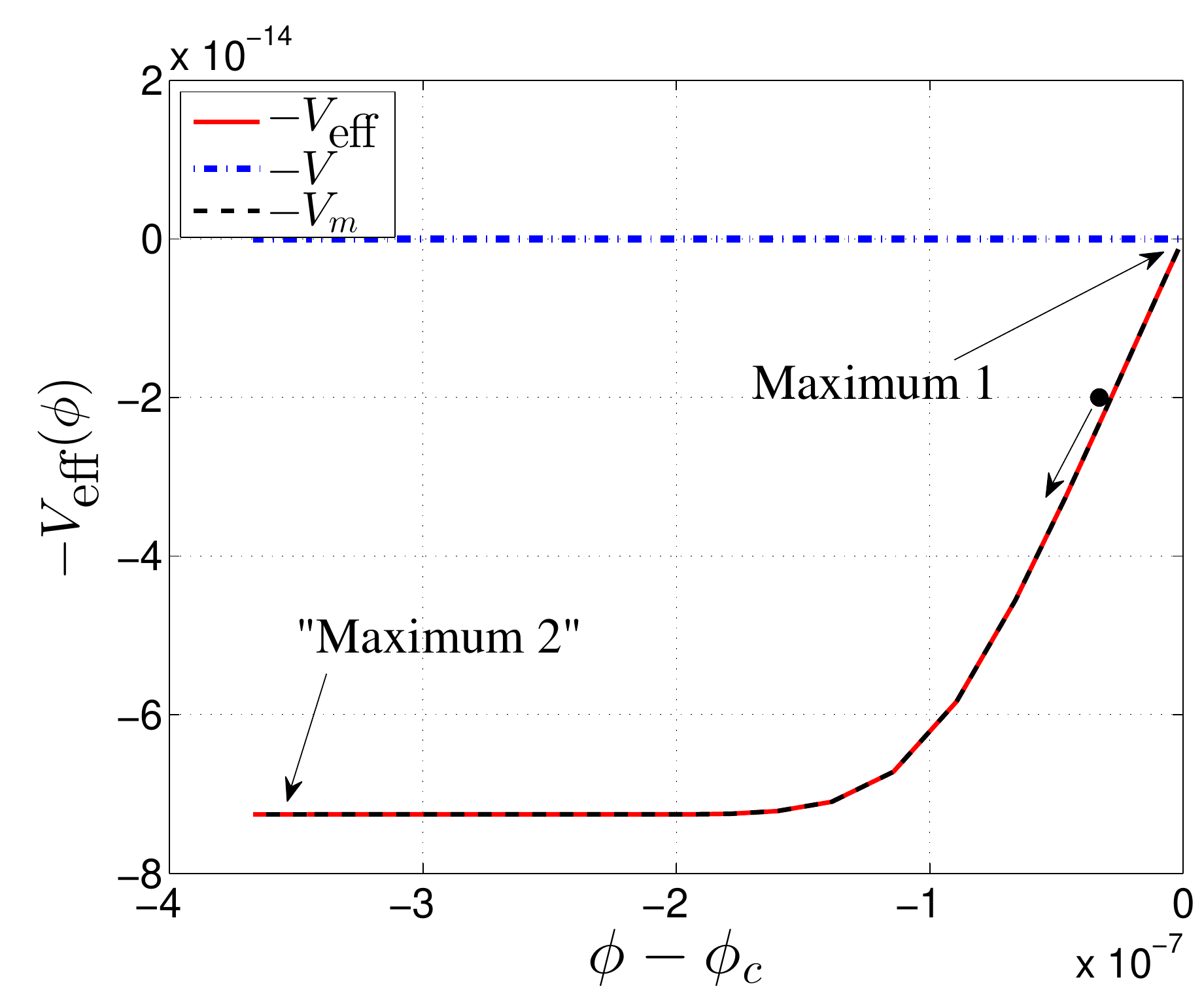, width=5.9cm} \\
    \hspace{10pt}(c)\\
  \end{tabular}
  \caption{(Color online) Numerical solution for the $R\ln R$ model in the solar system case
  (a) The field $\phi(r)$. In this configuration, $\phi(r)$ is not coupled to the matter density inside the Sun, and then a shell does not exist near the surface of the Sun. $\phi_c(=0.202555860)$ is the value of $\phi$ at the center of the Sun.
  (b) The terms in the equation of motion for $\phi$ (\ref{approx_trace_eq}). The friction force $2\phi'/r$ is relatively large, and $\phi''+2\phi'/r \approx 8\pi G\rho/3$, which leads to a metric Eq.~(\ref{metric_f_R}) different from the observations.
  (c) The inverted effective potential $-V_\text{eff}(\phi)$. The field $\phi$ slowly rolls down from Maximum 1 to \lq\lq Maximum 2\rq\rq~due to the large friction force. $V_{\text{eff}}=V+V_m$.
  }
  \label{fig:solution_RlnR_phi_light}
\end{figure*}

In the above two configurations, inside the sphere, $\phi$ stays at the coupling state. Then, the scalar field $\phi$ runs slowly with respect to the matter density, see Eq.~(\ref{phi_RlnR}). This running can easily trigger the field $\phi$ to move off the coupling state inside the sphere and then the field $\phi$ will slowly approach the other coupling state which is far away from the sphere, as shown in Figs.~\ref{fig:solution_RlnR_set2} and \ref{fig:solution_RlnR_set1}. The smaller the matter density and the radius of the sphere are, the earlier the field $\phi$ will be released from the coupling state inside the sphere. In the solar system case, such a coupling process does not even exist. We let $\alpha_0=0.0126$ and $R_0=1$ so that $\Lambda$ can take the value of the dark energy, $10^{-34}$. The numerical results in the solar case are shown in Fig.~\ref{fig:solution_RlnR_phi_light}. Figure~\ref{fig:solution_RlnR_phi_light}.a shows that, Eq.~(\ref{approx_trace_eq}) can be approximated as $\phi''+2\phi'/r \approx 8\pi G\rho/3$ and $\phi''+2\phi'/r \approx 0$ inside and outside of the Sun, respectively. As a result, outside of the Sun, $\phi(r) \approx \phi_0 +2GM/(3r)$, where $\phi_0=2\alpha_0 + \alpha_0 W(8\pi G\rho_{\text{\tiny{solar-system}}}/\Lambda)$. Thus, a metric close to Eq.~(\ref{metric_f_R}), which is different from the observations, will be obtained, like what has been discussed in Sec.~\ref{sec:gamma_1_over_2}.

One may also interpret the results in the solar case in the other way. For the $R\ln R$ model,
\be f''=\frac{\alpha_0}{R}. \label{f_dubprime_RlnR}\ee
In comparison to the requirement of the general relativity restoration (\ref{condition_on_f_R_2}), the modification term in the $R\ln R$ model changes fast with respect to the Ricci scalar $R$, and the model deviates significantly from general relativity. Consequently, with Eq.~(\ref{phi_RlnR}), there can be a big gap for $\phi$ between inside and outside of the sphere. In the solar system, at the outside of the Sun, due to the small value of $W(8\pi G\rho_{\text{solar-system}}/\Lambda)$,
the value of $\phi$ in the coupling state, $V'(\phi)\approx 8\pi G\rho_{\text{solar-system}}/3$, is very close to its de Sitter value $2\alpha_{0}$ corresponding the minimum of the potential $V(\phi)$. Then along the radial direction from outside to inside the Sun, because of the big gap between the general relativity restoration value $\phi\approx 1$ and the value close to $\phi_0(=2\alpha_0)$, $\phi$ slowly moves toward the general relativity restoration place as the matter density increases. However, because the solar density is not high enough and the Sun is not large enough for the $R\ln R$ model, inside the Sun, the equation of motion for $\phi$ is reduced to $\phi''+2\phi'/r \approx 8\pi G\rho_{\text{Sun}}/3$. In other words, the field $\phi$ even does not come to the equilibrium point between $V'(\phi)$ and $8\pi G\rho_{\text{Sun}}/3$ as the radius approaches to zero. As a result, Eq.~(\ref{v_dub_prime_matter_domi}) does not apply, and $\phi(r)$ does not have a shell across the surface of the Sun.

As a supplement, Fig.~\ref{fig:solution_RlnR_phi_light}.c shows that the inverted potential $-V_{\text{eff}}$ only has one rather than two maxima, so $\phi$ just slowly rather than instantly rolls down from the maximum (inside the Sun) to the minimum (outside the Sun) of the inverted potential $-V_{\text{eff}}$.  In summary, due to the large running of the modification term with respect to the Ricci scalar $R$, the $R\ln R$ model has difficulty to pass the solar system tests.

As a result, it remains challenging for the $R\ln R$ model to pass the solar system tests when the chameleon mechanism has been taken into account. This problem is significantly alleviated in some other $f(R)$ models (e.g. the Hu-Sawicki model) that are close to the $\Lambda$CDM model. In these models, the field $\phi$ is not sensitive to the change of the matter density when the matter density is above the cosmological constant scale.
\subsection{The Hu-Sawicki model}
The function $f(R)$ in the Hu-Sawicki model reads~\cite{Wayne_Hu}
\be f(R) = R - R_0\frac{C_1R^{n}}{C_2R^{n}+R_{0}^{n}},\label{Hu_Sawicki_model}\ee
where $C_1$ and $C_2$ are dimensionless parameters, $R_0=8\pi G\bar{\rho}_0/3$, and $\bar{\rho}_0$ is the average matter density of the current universe.
We consider the simplest version of this model
\be f(R) = R-\frac{CR_0 R}{R+R_0}, \ee
where $C$ is a dimensionless parameter. In this model,
\be V'(\phi)=\frac{R^{3}}{3(R+R_0)^{2}}\left[1+(1-C)\frac{R_0}{R}\left(2+\frac{R_0}{R}\right) \right]. \label{V_prime_Hu_Sawicki}\ee
The above two equations show that as long as the matter density is much greater than $R_0$, the curvature $R$ will trace the matter density well,
and $\phi$ will be close to $1$.
\begin{figure*}[t!]
  \begin{tabular}{ccc}
    \hspace{-5pt}\epsfig{file=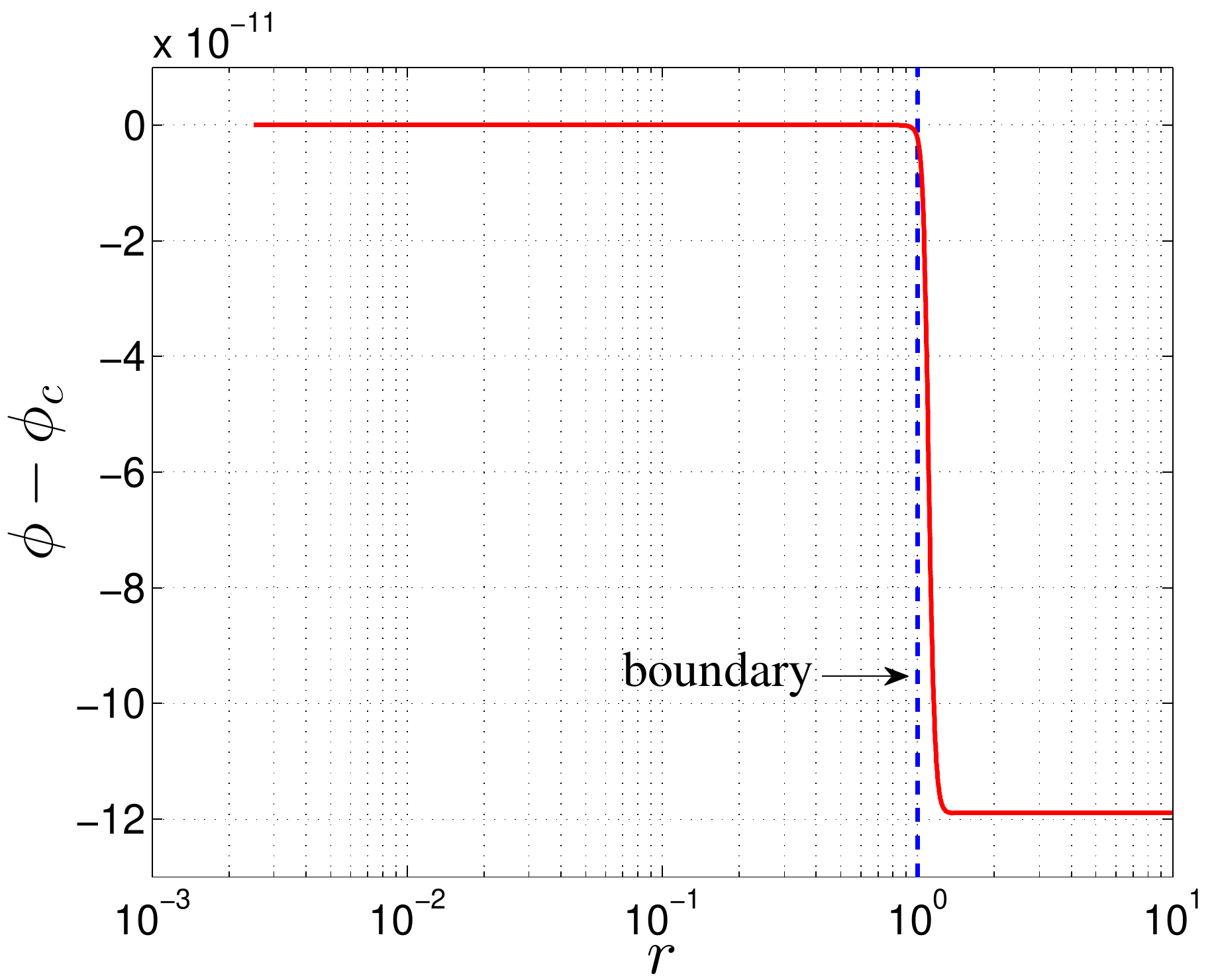, width=6cm,height=4.8cm} &
    \hspace{5pt}\epsfig{file=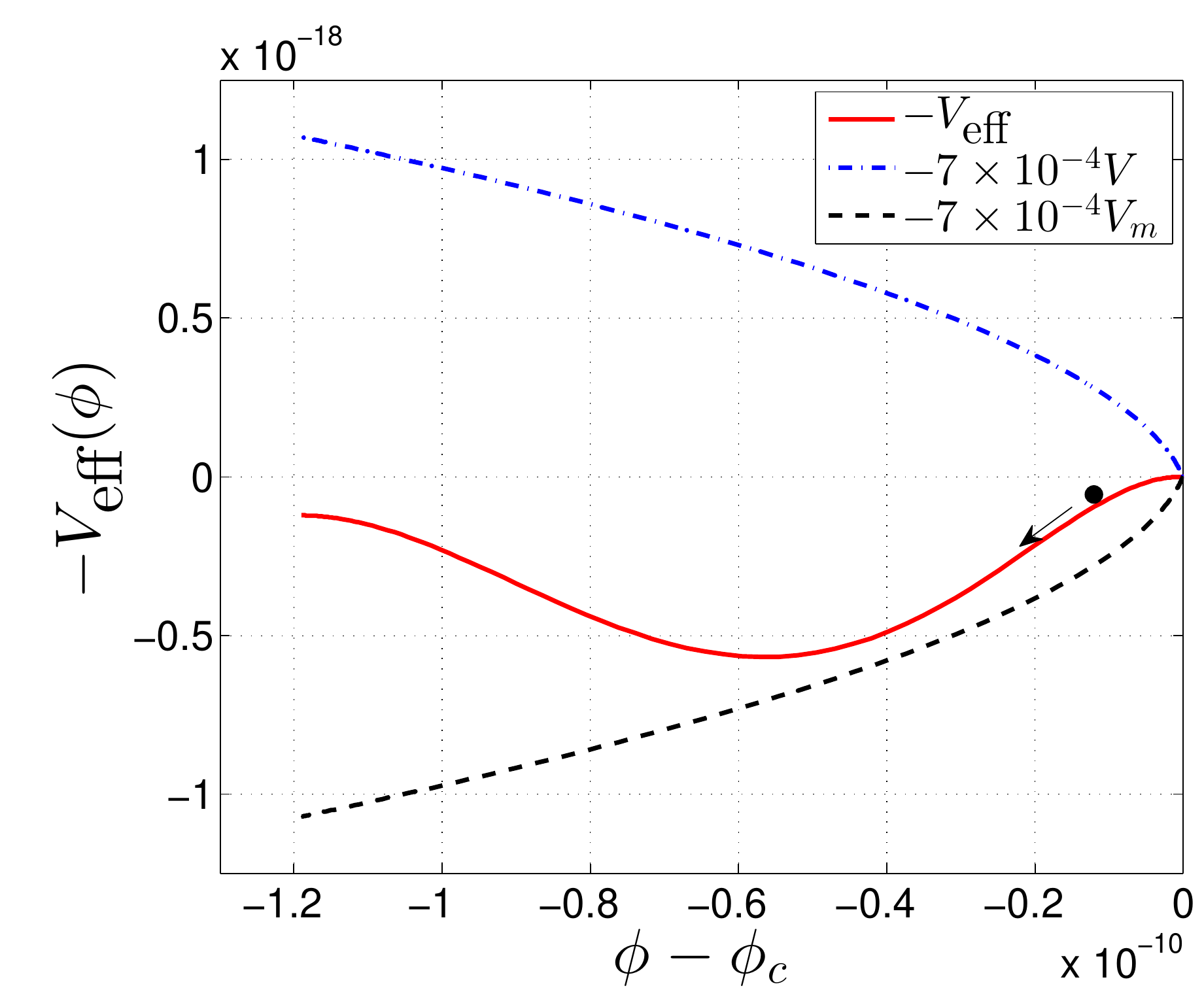, width=6.3cm,height=4.8cm} \\
    \hspace{10pt}(a) &
    \hspace{32pt}(b) \\
  \end{tabular}
  \caption{(Color online) Numerical solution for the Hu-Sawicki model.
  (a) The field $\phi(r)$. A thin shell exists near the surface of the sphere.
   $\phi_{c}(=1-1.188\times10^{-14})$ is the value of $\phi$ at the center of the sphere.
  (b) The inverted effective potential $-V_\text{eff}(\phi)$. The two maxima of $-V_\text{eff}(\phi)$ are almost at the same height, and $\phi(r)$ is an instanton in $-V_\text{eff}(\phi)$. $V_{\text{eff}}=V+V_m$.
  }
  \label{fig:solution_Hu_Sawicki}
\end{figure*}

In this model,
\be A''(R)=\frac{2CR_0}{(R+R_0)^{3}}\ll \frac{1}{R}, \mbox{ when $R\gg R_0$}\nonumber \ee
where $A(R)$ is the modification term, $f(R)=R+A(R)$. So $A'(R)$ moves very slowly with respect to $R$ in comparison to $1/R$, which makes the Hu-Samicki model favorable to avoid the solar system tests. We compute the field $\phi(r)$ when a sphere sits in a background with nonzero matter density. The parameters assume the following values. Equation~(\ref{V_prime_Hu_Sawicki}) shows that in order for the model to have a de Sitter point for which $V'(\phi)=0$, the parameter $C$ needs to be greater than 1. In this paper, we set $C$ to be equal to $1.2$. In the same set of units as used in the $R\ln R$ case, the radius of the sphere $r_0$ is equal to 1, and the densities of the sphere, background, and $R_0$ are $2.1\times 10^{-4}$, $10^{-2}\rho_{\text{sphere}}$, and $10^{-7}\rho_{\text{sphere}}$, respectively. The density profile is $\rho = \rho_{\text{sphere}}/[1+\exp(30(r-r_0))] + \rho_{\text{background}}$.
The solution for $\phi(r)$ is plotted in Fig.~\ref{fig:solution_Hu_Sawicki}.a, which shows that $\phi(r)$ has a thin shell near
the surface of the sphere. As shown in Fig.~\ref{fig:solution_Hu_Sawicki}, the two maxima of the inverted effective potential $-V_{\text{eff}(\phi)}$ are at the same height, and $\phi(r)$ is an instanton in $-V_{\text{eff}(\phi)}$.

In principle, when the computational precision is high enough, numerical computations in the solar case can be performed. On the other hand, in this study we only considered a simple case ($n$ is equal to 1) of the Hu-Sawicki model (\ref{Hu_Sawicki_model}). The model with a larger $n$ will be more favored to pass the solar system tests because the field $\phi$ will be suppressed more in this circumstance. However, implementing all such computations is beyond the scope of this paper.

\vspace{-0.3cm}
\section{Conclusions\label{sec:conclusions}}
The confrontation between $f(R)$ gravity and the solar system tests has been explored in the Jordan frame in this paper. The metric is in a gross violation of the observation if the Sun is assumed to sit in a vacuum background. We reobtain this result in a simpler way by directly focusing on the equation of motion for $\phi$ in the Jordan frame.

The chameleon mechanism implies that the functional form $f(R)$ should be very close to the Ricci scalar $R$ for $R$ above or equal to the solar system scale,
\be |A(R)|\ll R, \mbox{ } |A'(R)|\ll 1, \mbox{ and } A''(R)\ll 1/R, \nonumber \ee
where $A(R)$ is a modification term, $f(R)=R+A(R)$. On the other hand, the $f(R)$ gravity should deviate from general relativity at the cosmological scale, so that a cosmic acceleration can be generated in the late universe. Therefore, a trade-off between the requirements at high and low curvature scales needs to be made.

We numerically compute the profiles of $\phi(r)$ for a sphere (which can be the Sun) in an environment (which can be the solar system) with non-zero matter density in difference configurations, and the corresponding inverted effective potentials are plotted. These provide an intuitive approach to understand the effects from the matter density. Regarding the $R\ln R$ model, in the coupling state the scalar field $\phi$ runs logarithmically with respect to the matter density when
$8\pi G\rho/\Lambda$ is much greater than 1. This relationship can easily trigger the field $\phi$ to release from the coupling state inside the sphere, and $\phi$ will then slowly approach the other coupling state, which is far away from the sphere. Consequently, it is challenging for the $R\ln R$ model to pass the solar system test, and in fact the coupling state even does not exist inside the Sun for this model. In some other $f(R)$ models which are very close to the $\Lambda$CDM model, the field $\phi$ is robust to the change of the matter density as long as the matter density is greater than the cosmological constant. As a result, this class of $f(R)$ models have the advantage of passing the solar system tests.

\bigskip

\section*{Acknowledgments}\small
The author would  like  to thank Christopher P. L. Berry, Andrei V. Frolov, Aaron Plahn, Levon Pogosian, Howard Trottier, Shaojie Yin, and Chongming Xu  for useful discussions, and the University of Saskatchewan for its hospitality during the Black Holes IX conference. The author also thanks the referee for the helpful comments.
\\


\end{document}